\documentclass[sigconf, nonacm]{acmart}
\usepackage[conf={restate,no link to proof}]{proof-at-the-end}
\usepackage{comment}
\usepackage{amsthm}
\usepackage{xcolor}
\usepackage[linesnumbered,ruled,vlined]{algorithm2e}
\usepackage{etoolbox}
\usepackage{makecell}

\newbool{ARXIV}
\booltrue{ARXIV}
\usepackage{enumitem}

\usepackage{color} %
\usepackage{xargs} %
\usepackage{xspace} %

\def\notationcolor{black} %

\newcommand{\notation}[2]{\newcommand{#1}{{\textcolor{\notationcolor}{\ensuremath{#2}}}}}

\newcommand{\term}[2]{\newcommand{#1}{\textcolor{\notationcolor}{#2}\xspace}}

\notation{\data}{\mathcal{D}}
\notation{\mech}{\mathcal{M}} %
\notation{\datavec}{\vec{x}} %
\notation{\covar}{\mathbf{\Sigma}} %
\notation{\bmat}{\mathbf{B}} %
\notation{\comvar}{\covar_{*}} %
\notation{\comb}{\bmat_{*}} %
\notation{\commech}{\mech_{*}} %
\notation{\amat}{\mathbf{A}} %
\notation{\pcostmat}{\mathbf{X}}
\notation{\nullspace}{\mathbf{H}}
\notation{\dimsize}{d}
\notation{\query}{\vec{q}}
\notation{\qmat}{\mathbf{Q}}
\notation{\numquery}{m}
\notation{\senstwo}{\Delta_2}
\notation{\brank}{k}
\notation{\randalg}{\mathcal{A}}
\notation{\loewner}{\preceq}
\notation{\rloewner}{\succeq}
\notation{\rowspace}{\text{rowspace}}
\notation{\trace}{\text{trace}}

\notation{\threshold}{\mathbf{T}}
\notation{\outp}{\omega}

\term{\obj}{\textbf{OBJ}}

\term{\myalg}{CM}

\newtheoremstyle{mystyle}%
  {}%
  {}%
  {\itshape}%
  {}%
  {\scshape}%
  {.}%
  { }%
  {}%
\theoremstyle{mystyle}

\newtheorem{theorem}{Theorem}[section]

\newtheorem{definition}[theorem]{Definition}

\newtheorem{lemma}[theorem]{Lemma}
\newtheorem{remark}[theorem]{Remark}

\let\vec\mathbf  %
\let\mat\mathbf  %

\AtBeginDocument{%
  \providecommand\BibTeX{{%
    \normalfont B\kern-0.5em{\scshape i\kern-0.25em b}\kern-0.8em\TeX}}}

\def\asfname{$\%\mech_1$\xspace}
\def\ascname{$\%\mech_2$\xspace}

\begin{document}
\title{Answering Private Linear Queries Adaptively using the Common Mechanism}

\newcommand\vldbdoi{}
\newcommand\vldbpages{}
\newcommand\vldbvolume{}
\newcommand\vldbissue{}
\newcommand\vldbyear{}
\newcommand\vldbauthors{\authors}
\newcommand\vldbtitle{\shorttitle} 
\newcommand\vldbavailabilityurl{https://github.com/cmla-psu/commonmech}
\newcommand\vldbpagestyle{plain}

\author{Yingtai Xiao$^1$, Guanhong Wang$^2$, Danfeng Zhang$^1$, Daniel Kifer$^1$} 
\affiliation{%
  \institution{$^1$Penn State, $^2$University of Maryland, College Park}
}
\email{yxx5224@psu.edu,   guanhong@umd.edu,    dbz5017@psu.edu,    duk17@psu.edu}

\begin{abstract}
    When analyzing confidential data through a privacy filter, a data scientist often needs to decide which queries will best support their intended analysis. For example, an analyst may wish to study noisy two-way marginals in a dataset produced by a mechanism $\mech_1$. But, if the data are relatively sparse, the analyst may choose to examine noisy one-way marginals, produced by a mechanism $\mech_2$, instead. Since the choice of whether to use $\mech_1$ or $\mech_2$ is data-dependent, a typical differentially private workflow is to first split the privacy loss budget $\rho$ into two parts: $\rho_1$ and $\rho_2$, then use the first part $\rho_1$ to determine which mechanism to use, and the remainder $\rho_2$ to obtain noisy answers from the chosen mechanism. In a sense, the first step seems wasteful because it takes away part of the privacy loss budget that could have been used to make the query answers more accurate.

In this paper, we consider the question of whether the choice between $\mech_1$ and $\mech_2$ can be performed without wasting any privacy loss budget. For linear queries, we propose a method for decomposing $\mech_1$ and $\mech_2$ into three parts: (1) a mechanism $\commech$ that captures their shared information, (2) a mechanism $\mech_1^\prime$ that captures information that is specific to $\mech_1$, (3) a mechanism $\mech_2^\prime$ that captures information that is specific to $\mech_2$. Running $\commech$ and $\mech_1^\prime$ together is completely equivalent to running $\mech_1$ (both in terms of query answer accuracy and total privacy cost $\rho$). Similarly, running $\commech$ and $\mech_2^\prime$ together is completely equivalent to running $\mech_2$.

Since $\commech$ will be used no matter what, the analyst can use its output to decide whether to subsequently run $\mech_1^\prime$ (thus recreating the analysis supported by $\mech_1$) or $\mech_2^\prime$ (recreating the analysis supported by $\mech_2$), without wasting privacy loss budget.
\end{abstract}

\maketitle

\pagestyle{\vldbpagestyle}
\begingroup\small\noindent\raggedright\textbf{PVLDB Reference Format:}\\
\vldbauthors. \vldbtitle. PVLDB, \vldbvolume(\vldbissue): \vldbpages, \vldbyear.\\
\href{https://doi.org/\vldbdoi}{doi:\vldbdoi}
\endgroup
\begingroup
\renewcommand\thefootnote{}\footnote{\noindent
This work is licensed under the Creative Commons BY-NC-ND 4.0 International License. Visit \url{https://creativecommons.org/licenses/by-nc-nd/4.0/} to view a copy of this license. For any use beyond those covered by this license, obtain permission by emailing \href{mailto:info@vldb.org}{info@vldb.org}. Copyright is held by the owner/author(s). Publication rights licensed to the VLDB Endowment. \\
\raggedright Proceedings of the VLDB Endowment, Vol. \vldbvolume, No. \vldbissue\ %
ISSN 2150-8097. \\
\href{https://doi.org/\vldbdoi}{doi:\vldbdoi} \\
}\addtocounter{footnote}{-1}\endgroup

\ifdefempty{\vldbavailabilityurl}{}{
\vspace{.3cm}
\begingroup\small\noindent\raggedright\textbf{PVLDB Artifact Availability:}\\
The source code, data, and/or other artifacts have been made available at \url{\vldbavailabilityurl}.
\endgroup
}

\section{Introduction}\label{sec:intro}
Consider an analyst who is working with confidential demographics data through a differential privacy filter -- the analyst poses queries and receives noisy answers. The analyst has a privacy loss budget $\rho$ and wishes to use it to study age and race distributions in a region. The analyst wishes to get a noisy age by race marginal (115 age categories and 63 race categories based on the latest census, for a total of $63\times 115$ cells). However, for small populations, this marginal would be sparse and the noise would swamp the data. In that case, the analyst could prefer two one-way marginals: one marginal on race and a separate marginal on age. The analyst does not know in advance whether the two-way race by age marginal  (Option 1) is better for this region or if two one-way marginals  (Option 2) are better.

In a typical workflow, the analyst would split the privacy loss budget $\rho$ into two pieces $\rho_1$ and $\rho_2$ (with $\rho_1+\rho_2=\rho$). The first piece would be used to somehow determine which of the two  options would provide a good signal to (privacy) noise ratio. For example, the analyst could ask for a noisy population total to make the decision, or could use the exponential mechanism \cite{exponentialMechanism}, which is a common technique for selecting among several options.
The remaining privacy loss budget $\rho_2$ would be used to provide a noisy answer to the chosen option (either a noisy age by race marginal, or to the two one-way marginals).

Now, this procedure comes with some regret because, if the analyst had known in advance which option to pick, then the entire privacy loss budget $\rho$ (instead of only $\rho_2$) could have been used to provide a noisy marginal, providing more accuracy. Thus the analyst may feel that $\rho_1$, the portion of the privacy loss budget used to select between the two options, was wasted or lost.

In this paper, we consider the question of how the analyst can choose between Options 1 and 2 so that no privacy loss budget is lost, and the entire $\rho$ is spent on the chosen analysis. Suppose $\mech_1$ is the mechanism used to provide noisy answers in Option 1 with privacy budget  $\rho$ and $\mech_2$ is the mechanism used to provide noisy answers in Option 2 with privacy budget $\rho$. We show how to split $\mech_1$ into two mechanisms $\commech$ and $\mech_1^\prime$, so that running $\mech_1$ with privacy budget $\rho$ is completely equivalent to running $\commech$ and $\mech_1^\prime$ together. Similarly, we split $\mech_2$ into $\commech$ and $\mech_2^\prime$. This  ``common'' mechanism $\commech$ represents information that is common to both $\mech_1$ and $\mech_2$. That is, this is a piece of information that would be provided to the analyst by either mechanism. Meanwhile, the ``residual'' $\mech_1^\prime$ encapsulates information that is specific to $\mech_1$ (one can think of $\mech_1^\prime$ as the result of removing the information in $\commech$ from $\mech_1$). Similarly, $\mech_2^\prime$ is the information that is specific to $\mech_2$.
The analyst's workflow becomes the following.
\begin{itemize}[leftmargin=*]
\item[1.] Given a dataset $\data$,  first run $\commech(\data)$ to get an output $\outp_*$. 
\item[2a.] Based on $\outp_*$, the analyst can choose to run the residual mechanism $\mech^\prime_1(\data)$ to get an output $\outp^\prime_1$, 
\item[2b.] Or, based on $\outp_*$, the analyst can instead run the residual mechanism $\mech^\prime_2(\data)$ to get an output $\outp_2^{\prime}$.
\end{itemize}
If the analyst runs $\commech$ followed by $\mech_1^\prime$, the total privacy budget spent is $\rho$ and the resulting outputs $\outp_*$ and $\outp^\prime_1$ provide the same information as if $\mech_1$ had been run with privacy budget $\rho$. Similarly, if the analyst runs $\commech$ followed by $\mech_2^\prime$, then the total privacy budget spent again is $\rho$ and the resulting outputs $\outp_*$ and $\outp^\prime_2$ provide the same information as running $\mech_2$ with privacy budget $\rho$. 
Thus, the analyst adaptively chooses which mechanism to run without any wasted privacy budget.

Our technique works for any mechanism that answers linear queries with Gaussian noise. It is compatible with Renyi differential privacy \cite{renyidp}, zCDP \cite{zcdp}, Gaussian Differential Privacy \cite{fdp},  $(\epsilon,\delta)$-differential privacy \cite{dworkKMM06:ourdata}, and personal differential privacy \cite{personaldp}. 

This kind of scenario, where an analyst needs to choose between pre-specified analyses depending on which provides an appropriate signal to (privacy) noise ratio, is expected to become more common. For example, the 2020 Census Detailed Demographic and Housing Characteristics data products are going to include sex-by-age tabulations where the binning of age in a region is data-dependent \cite{dhc}. For small regions, only the population totals will be published. For more populous regions, the age will be binned into 4, 9, or 23 buckets, depending on how populous the region is. To choose which bucketization to use, noise will be first added to the population in a region \cite{dhc}. This noisy count will be checked against manually-specified thresholds to determine the buckets to use. We empirically show that our proposed approach is better at selecting the correct analysis, removes the need for manual tuning, and uses all of the privacy loss budget on the sex-by-age histograms instead of taking away some of it for the purposes of selecting the histogram to use.

The contributions of this paper are the following: 
\begin{itemize}[leftmargin=*,topsep=4pt]
\item We provide a framework for adaptively choosing between two linear mechanisms $\mech_1$ and $\mech_2$ without requiring additional expenditure of privacy loss budget.
\item We formalize the definition of the common mechanism $\commech$ of $\mech_1$ and $\mech_2$ as an optimization problem, and also formalize the associated residual mechanisms $\mech_1^\prime,\mech_2^\prime$. This framework can be extended to the case of multiple mechanisms (e.g., $\mech_1,\mech_2,\mech_3,\dots$), but the optimization problem for the common mechanism has an analytical solution when dealing with two mechanisms.
\item We provide algorithms for computing the common and residual mechanisms.
\item We give suggestions on how to use the output of the common mechanism to decide between $\mech_1$ and $\mech_2$, thus providing another tool for the construction of differentially private systems.
\item We demonstrate the efficacy of this approach using real datasets and apply it to a real-world application involving census data.
\end{itemize}

We present notation and background in Section \ref{sec:background}. We formally define the problem statement in Section \ref{sec:problemdef}. As we believe our techniques can have wider applicability than what we can cover here, we also discuss some possibilities of future work in Section \ref{sec:problemdef} as well. Related work is  discussed in Section \ref{sec:related}. Algorithms for the common and residual mechanisms are in Section \ref{sec:common}. Suggestions on how to decide between mechanisms based on the output of the common mechanism are in Section \ref{sec:decide}. Experiments are in Section \ref{sec:experiments} and conclusions are in Section \ref{sec:conc}. All proofs can be found in the full version of this paper at \url{https://github.com/cmla-psu/commonmech}.

\section{Notation and Background}\label{sec:background}

In this section, we explain our notation (summarized  in Table \ref{tab:notation}) and provide background information on differential privacy and the type of mechanisms we consider. 

We denote vectors as bold lower-case letters (e.g., $\datavec$), matrices as bold upper-case (e.g., $\bmat$), scalars as non-bold lower-case (e.g., $\sigma$). If $\mat{A}$ and $\mat{B}$ are positive semidefinite matrices, we say $\mat{B}\loewner \mat{A}$ if $\mat{A}-\mat{B}$ is positive semidefinite ($\rloewner$ is defined analogously). The relation $\loewner$ defines a partial ordering on semidefinite matrices called the \emph{Loewner} order \cite{horn2012matrix}.

A \textbf{dataset} $\data$ is a table of records.
Following earlier work on differentially private linear queries \cite{YYZH16,LMHMR15,YZWXYH12,xiao2020optimizing}, we assume the record attributes are categorical (or have been discretized). As in prior work, we represent the dataset $\data$  as a vector $\datavec$ of counts and we refer to it as the \textbf{data vector}. That is, letting $\{t_0, t_1, \dots t_{\dimsize-1}\}$ be the set of possible record values,  $\datavec[i]$ is the number of times record $t_i$ appears  in  $\data$.
For example, if each record consists of two attributes, \textit{adult} (yes/no) and \textit{Hispanic} (yes/no), then there are 4 possible types of records, which are  $t_0 = $ "not adult, not Hispanic", $t_1=$``adult, not Hispanic'', $t_2=$``not adult, Hispanic'', $t_3=$``adult, Hispanic''. In our representation, $\datavec[3]$ is the number of Hispanic adults in the dataset $\data$. 

We say that two dataset vectors $\datavec$ and $\datavec^\prime$ are \textbf{neighboring} (denoted as $\datavec \sim \datavec^\prime$) if $\datavec$ can be obtained from $\datavec^\prime$ by adding or subtracting 1 from some component of $\datavec^\prime$ (this means $||\datavec-\datavec^\prime||_1=1$) -- this is the same as adding or removing 1 person from the underlying dataset.

A single \textbf{linear query} $\query$ is a vector, whose answer is $\query\cdot \datavec$. 
A \textbf{query set} is a set of $\numquery$ linear queries represented by an $\numquery\times \dimsize$ matrix $\bmat$, where each row corresponds to a single linear query. We let $\brank$ denote the rank of $\bmat$. The answers to the queries are  obtained by matrix multiplication: $\bmat \datavec$.
Continuing our running example of a two-attribute  dataset, if $ \bmat =\left(\begin{smallmatrix}0 & 1 & 0 & 1\\1 & 1 & 1 & 1\end{smallmatrix}\right)$, then this is a set of two queries. The first query is the  number of times records $t_1$ or $t_3$ appear in the dataset (i.e., the number of adults) and the second query is the total number of people.

A \textbf{mechanism} $\mech$ is an algorithm whose input is the confidential data (either $\data$ or $\datavec$) and whose output $\outp$ is considered safe to release (because it protects privacy). When we have two mechanisms $\mech_a$ and $\mech_b$, we let $(\mech_a, \mech_b)$ denote the mechanism that runs both of them on the data and releases their results. In other words, its output is $(\mech_a(\data), \mech_b(\data))$.

\subsection{Differential Privacy}

Differential Privacy \cite{dwork06Calibrating,dworkKMM06:ourdata,zcdp,renyidp,fdp} is a family of privacy definitions that place restrictions on how a mechanism $\mech$ can work. It has become a de facto standard for protecting confidentiality when creating publicly available data products, with an ever-increasing list of real-world deployments, including the U.S. Census Bureau \cite{ashwin08:map,onthemap,tdahdsr}, Uber \cite{elasticsensitivity,chorus}, Apple \cite{appledpscale}, Facebook \cite{fburlshares,opacus}, Microsoft \cite{DingKY17}, and Google \cite{rappor,tensorflowprivacy,googlesql}. Differential privacy provides a rigorous plausible deniability guarantee -- it limits the ability of an attacker to determine whether a target person's record was in the dataset or not. The most common variation of differential privacy is:

\begin{definition}[Approximate Differential Privacy \cite{dworkKMM06:ourdata}]\label{def:dp} Given privacy parameters $\epsilon > 0$ and $\delta \in (0, 1)$, a randomized algorithm $\mech$ satisfies $(\epsilon,\delta)$-differential privacy if for all pairs of neighboring dataset vectors $\datavec$ and $\datavec^\prime$ and all sets $S$, the following equations hold:
\begin{align*}
    P(\mech(\datavec)\in S) \leq e^{\epsilon} P(\mech(\datavec^\prime)\in S) + \delta
\end{align*}
\end{definition}

\begin{figure}
    \centering
    \includegraphics[width=0.4\textwidth,clip=true,trim=0.4cm 0.6cm 0cm 0cm]{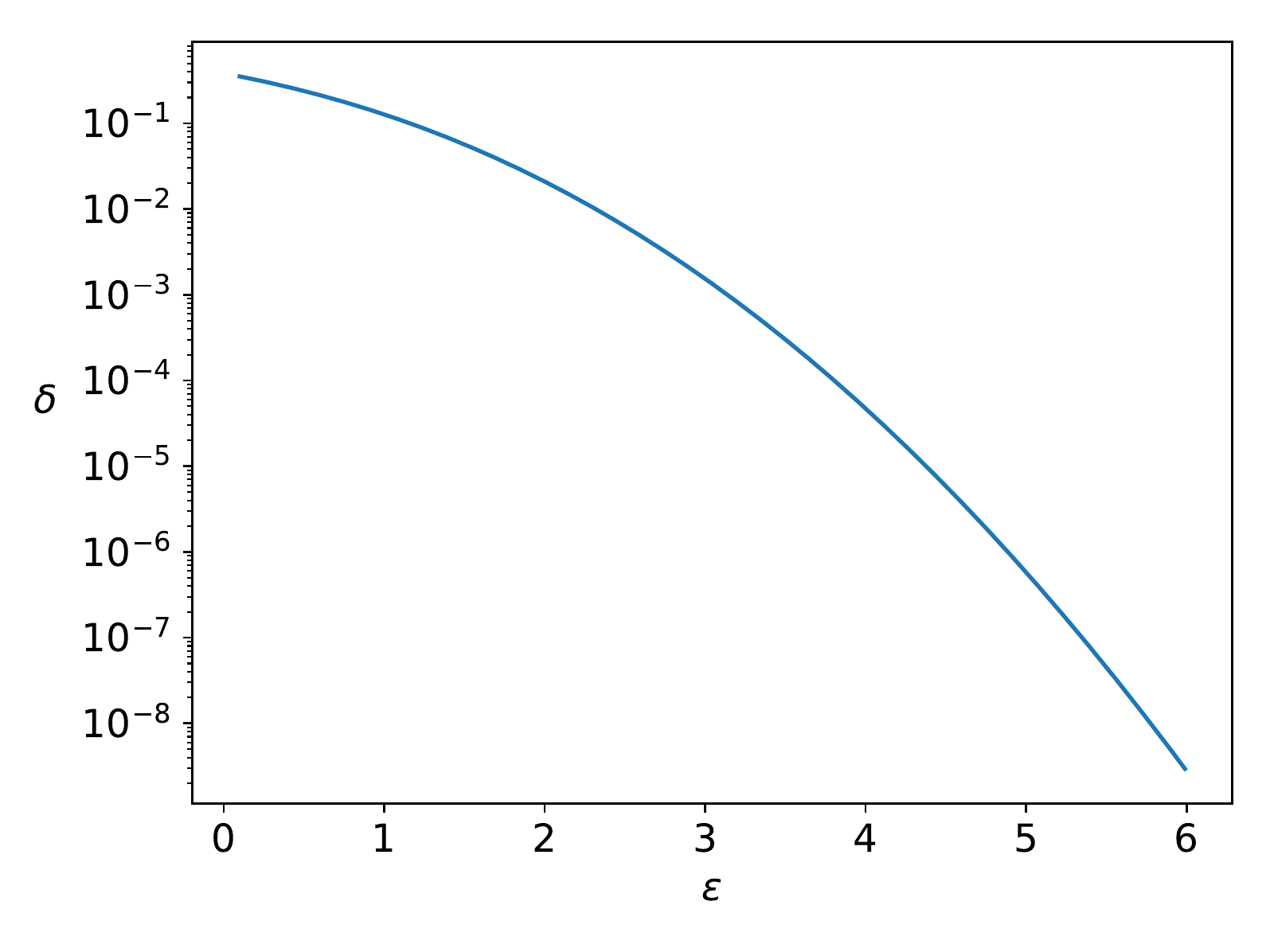}
    \caption{Example of an $(\epsilon,\delta)$ curve for the mechanism $\mech(\datavec)=\vec{1}^T\datavec + N(0, 1)$.}
    \label{fig:curve}
\end{figure}

A mechanism $M$ typically satisfies approximate differential privacy for infinitely many $(\epsilon, \delta)$ pairs, which defines a curve in space. An example curve is shown in  Figure \ref{fig:curve}.
There is also another popular variant known as \emph{zero-concentrated differential privacy} (zCDP):

\begin{definition}[zCDP \cite{zcdp}]\label{def:zcdp} Given a privacy parameters $\rho>0$, a randomized algorithm $\mech$ satisfies $\rho$-zCDP if for all pairs of neighboring dataset vectors $\datavec$ and $\datavec^\prime$ and all numbers $\alpha>1$, 
\begin{align*}
    \int_\outp P(\mech(\datavec)=\outp) \frac{P(\mech(\datavec)=\outp)^{\alpha-1}}{P(\mech(\datavec^\prime)=\outp)^{\alpha-1}}~d\outp \leq e^{(\alpha-1)\alpha \rho},
\end{align*}
where $P(\mech(\datavec)=\outp)$ is interpreted as a probability density function in the continuous case.
\end{definition}

The parameters $(\epsilon,\delta)$ are known as the \emph{privacy loss budget} of approximate differential privacy and the privacy parameter $\rho$ is known as the privacy loss budget of zCDP. Note that the privacy parameter of zCDP is difficult to interpret (see \cite{semantics} for an extended discussion) but easier to compute than approximate differential privacy. Thus, one typically determines the zCDP privacy loss parameter $\rho$ of a mechanism $\mech$  and then converts it to $\epsilon$ and $\delta$ for interpretability \cite{zcdp,discgauss}.

Each version of differential privacy also has a  ``personalized'' version, in which each possible record type $t_i$ is assigned a privacy loss budget. The privacy parameter for record type $t_i$ can be obtained from Definition \ref{def:dp}  or \ref{def:zcdp} by  considering only neighbors $\datavec$ and $\datavec^\prime$ that differ in their $i^\text{th}$ coordinate \cite{personaldp}.

\subsection{The Linear Gaussian Mechanism}\label{subsec:gaussmech}
The linear Gaussian mechanism adds Gaussian noise to the output of linear queries and is compatible with many versions of differential privacy. It is defined as follows.

\begin{definition}[Linear Gaussian Mechanism]\label{def:gaussmech} Given a \underline{query matrix} $\bmat$ and nonsingular \underline{covariance matrix} $\covar$ (not necessarily diagonal), the linear Gaussian mechanism $\mech$ is defined as $\mech(\datavec)=\bmat \datavec + N(\vec{0}, \covar)$. The quantity $\bmat^T \covar^{-1}\bmat$ is called  the \textbf{privacy cost matrix} of $\mech$.
\end{definition}

The importance of the privacy cost matrix is that the privacy parameters of the Gaussian mechanism for $(\epsilon,\delta)$-differential privacy and $\rho$-zCDP (both the basic and personalized versions) are all functions of this privacy cost matrix, as the following result shows:\footnote{This is also true of Renyi differential privacy \cite{renyidp} and Gaussian differential privacy \cite{fdp}.}

\begin{lemma}[\cite{xiao2020optimizing}]\label{corr:query}
Let $\mech$ be the linear Gaussian mechanism (Definition \ref{def:gaussmech}) defined as $\mech(\datavec)=\bmat \datavec + N(\vec{0},\covar)$ with privacy cost matrix $\mathbf{C}=\bmat^T \covar^{-1}\bmat$. Let $c_i$ be the $i^\text{th}$ diagonal entry of $\mathbf{C}$ and let $c_{max}=\max_i c_i$ be the largest diagonal entry of $\mathbf{C}$.  Let $\Phi$ be the CDF of the standard normal distribution. Then:
\begin{itemize}
\item $\mech$ satisfies $(\epsilon,\delta)$-differential privacy for $$\delta = \Phi\left(\frac{\sqrt{c_{max}}}{2}-\frac{\epsilon}{\sqrt{c_{max}}}\right) - e^{\epsilon}\Phi\left(-\frac{\sqrt{c_{max}}}{2}-\frac{\epsilon}{\sqrt{c_{max}}}\right)$$ and $\delta$ is an increasing function of $c_{max}$. In particular, this means the entire $(\epsilon,\delta)$ curve of $\mech$ is determined by $c_{max}$.
\item The personalized approximate differential privacy parameters $(\epsilon_i, \delta_i)$ for record type $t_i$ are obtained from the formula: $$\delta_i = \Phi\left(\frac{\sqrt{c_{i}}}{2}-\frac{\epsilon_i}{\sqrt{c_{i}}}\right) - e^{\epsilon_i}\Phi\left(-\frac{\sqrt{c_{i}}}{2}-\frac{\epsilon_i}{\sqrt{c_{i}}}\right)$$
\item $\mech$ satisfies $\rho$-zCDP for $\rho=c_{max}/2$.
\item The personal zCDP privacy parameter for record $t_i$ is $ c_i/2$.
\end{itemize}
\end{lemma}

\begin{table}[t]
\begin{center}
\caption{Table of Notation}\label{tab:notation}
\resizebox{1 \linewidth}{!}{
\begin{tabular}{|cl|}\hline
$\data$:& Dataset \\
$\datavec$: & Data vector representation of $\data$ \\
$\dimsize$:     & Number of possible records\\
$\mech$: & Mechanism. \\
$\outp$: & Output of a mechanism. \\
$(\mech_a, \mech_b)$ & Combined mechanism that runs $\mech_a$ and $\mech_b$\\
$\bmat$: & Query matrix.\\
$\numquery$: & Number of rows of $\bmat$ ($\bmat$ has size $\numquery\times \dimsize$).\\
$\brank$: & Rank of $\bmat$.\\
$\covar$: & Covariance matrix. \\
$\bmat^T\covar^{-1}\bmat$: & Privacy cost matrix of mechanism $\mech(\datavec)=\bmat\datavec + N(\vec{0}, \covar).$ \\
$\loewner, \rloewner$ & Loewner order ($\mat{A_2} \loewner\mat{A_2}$ iff $\mat{A_1}-\mat{A_2}$ is positive semidefinite)\\
\hline
\end{tabular}
}
\end{center}
\end{table}

\section{Problem Definition and Solution Overview}\label{sec:problemdef}
The motivation for our problem is the following. An analyst is interested in obtaining noisy linear query answers either from mechanism $\mech_1$ or $\mech_2$, defined as follows:
\begin{align*}
\mech_1(\datavec) &= \bmat_1 \datavec + N(0, \covar_1)\\
\mech_2(\datavec) &= \bmat_2 \datavec + N(0, \covar_2).
\end{align*}
where $\mech_1$ and $\mech_2$ both satisfy zCDP with the same privacy parameter $\rho$.\footnote{For concreteness, we focus on zCDP, but, as we show later, this approach works for any post-processing invariant privacy definition.}
For example, $\mech_1$ could compute all the one-way marginals of a dataset, and $\mech_2$ could compute all the two-way marginals. However, the choice of which mechanism to use depends on the properties of the data that the analyst does not know.

For instance, if the dataset is ``large enough'' then the noisy two-way marginals (i.e., $\mech_2$) will be very accurate with low relative error. Otherwise, the analyst would prefer to use the noisy one-way marginals produced by $\mech_1$. The problem is that the analyst does not know whether the dataset is ``large enough'', or even how to precisely define what ``large enough'' means (i.e., exactly how many records are needed for the dataset to be considered large enough). 

Thus, the analyst needs extra information about the data in order to make a choice between $\mech_1$ and $\mech_2$. One option is to take some privacy budget $\rho_0$ away from $\mech_1$ and $\mech_2$. This $\rho_0$ would be assigned to some other mechanism $\mech^\circ$ that queries that data. Based on the answers to $\mech^\circ$, the analyst would  modify either $\mech_1$ or $\mech_2$ to use the  remaining privacy budget $\rho-\rho_0$ (by rescaling the Gaussian covariance matrices $\covar_1$ and $\covar_2$) and then run it on the data. This option  produces noisier answers than the analyst wants, because only $\rho-\rho_0$ instead of $\rho$ is allocated towards the noisy answers. In this sense, this $\rho_0$ can feel like a wasted expenditure of privacy loss budget since it takes away from the accuracy of the query answers that the analyst desires.

In this paper, we consider a second option -- whether the analyst can make an informed choice without wasting any privacy loss budget. The main idea is to rewrite  mechanism $\mech_1$  as a sequence of two mechanisms $\commech$ and $\mech_1^\prime$ such that running $\mech_1$ is exactly equivalent, both in terms of information content and privacy cost,  to running $\commech$ and $\mech_1^\prime$ together. We similarly decompose $\mech_2$ into $\commech$ and $\mech_2^\prime$. This means that if the analyst wants to run $\mech_1$ or $\mech_2$, then $\commech$ is going to be run no matter what.
Thus the analyst can first run $\commech$ and then can decide whether to run $\mech_1^\prime$ (to finish the execution of $\mech_1$) or to run $\mech_2^\prime$ (to finish the execution of $\mech_2$).
We refer to $\mech_*$ as the mechanism that is \emph{common} to $\mech_1$ and $\mech_2$ since it represents the information that they share. We refer to $\mech_1^\prime$ and $\mech_2^\prime$ as the \emph{residual} mechanisms since they represent the information that is specific to $\mech_1$ and $\mech_2$, respectively.

We next formalize this discussion. We examine what it means for two mechanisms to be equivalent in Section \ref{subsec:equiv}, we formally define the common mechanism of $\mech_1$ and $\mech_2$ in Section \ref{subsec:shared}, we formally define the residual mechanisms in Section \ref{subsec:decompose}, and then we present the technical problem statement in Section \ref{subsec:statement}. We believe this technique may have wide applicability and discuss these possibilities, some of which are future work, in Section \ref{subsec:additional}.

\subsection{Answerability and Equivalence of Linear Gaussian Mechanisms}\label{subsec:equiv}

Suppose there are two mechanisms $\mech_a$ and $\mech_b$ and a randomized postprocessing algorithm $\randalg$ such that for all data vectors $\datavec$, the output distribution of $\mech_b(\datavec)$ is the same as the output distribution of $\randalg(\mech_a(\datavec))$. This means that we can simulate the output of $\mech_b$ by taking the output of $\mech_a$ and feeding it to $\randalg$. 

When this is the case, we say that $\mech_b$ is \emph{answerable} from $\mech_a$ and it means that $\mech_a$ produces at least as much information as $\mech_b$. For any post-processing invariant privacy definition (such as $\rho$-zCDP, Gaussian differential privacy, approximate differential privacy, etc.), the privacy cost of $\mech_a$ is also at least as large as the privacy cost of $\mech_b$ (e.g., the $\rho$ parameter of $\mech_a$ under zCDP is greater than or equal to the $\rho$ parameter of $\mech_b$).

For linear Gaussian mechanisms, answerability can be defined as follows.

\begin{definition}\label{def:answer}
Let $\mech_a(\datavec)=\bmat_a \datavec + N(0, \covar_a)$ and $\mech_b(\datavec)=\bmat_b\datavec + N(0, \covar_b)$ be linear Gaussian mechanisms. We say that $\mech_b$ is answerable from $\mech_a$ if there exist matrices $\mat{A}$ and $\mat{C}$ such that for every $\datavec$, $\mech_b(\datavec)$ has the same distribution as $\mat{A} \mech_a(\datavec) + \mat{C} N(0, \mat{I})$, where $\mat{I}$ is the identity matrix. In other words, $\mech_b$ can be obtained from $\mech_a$ by applying a linear transformation and adding additional noise (that does \textbf{not} depend on the data).
\end{definition}

For linear Gaussian mechanisms, answerability is easy to check using the following result whose proof is straightforward.

\begin{lemma}\label{lem:answer}
Let $\mech_a(\datavec)=\bmat_a \datavec + N(0, \covar_a)$ and $\mech_b(\datavec)=\bmat_b\datavec + N(0, \covar_b)$ be linear Gaussian mechanisms. $\mech_b$ is answerable from $\mech_a$ if and only if there exist matrices $\mat{A}$ and $\mat{C}$ such that:
\begin{align*}
\bmat_b &= \mat{A} \bmat_a\\
\covar_b &= \mat{A} \covar_a \mat{A}^T + \mat{C}\mat{C}^T
\end{align*}
\end{lemma}

\begin{example}\label{ex:answer}
Suppose $\datavec$ is two-dimensional and consider the mechanisms:
\begin{align*}
\mech_a(\datavec) &=\left[\begin{smallmatrix}1 & 1\\ 1 & 0\\
0 & 1\end{smallmatrix}\right] \datavec + N\left(\left[\begin{smallmatrix}
0\\0\\0\end{smallmatrix}\right], \left[\begin{smallmatrix}
2 & 0 & 0\\ 0 & 2 & 0\\ 0 & 0 & 2\end{smallmatrix}\right]\right)\\
\mech_b(\datavec) &=\left[\begin{smallmatrix} 1 & 0\\
0 & 1\end{smallmatrix}\right] \datavec + N\left(\left[\begin{smallmatrix}
0\\0\end{smallmatrix}\right], \left[\begin{smallmatrix}
4/3 & -2/3 \\ -2/3 & 4/3\end{smallmatrix}\right]\right)\\
\end{align*}
Both mechanisms satisfy $\rho$-zCDP with $\rho=1/2$.
For any $\datavec$, the output distribution of $\mech_b$ is the multivariate Gaussian with mean $\datavec$ and covariance matrix $\left[\begin{smallmatrix}
4/3 & -2/3 \\ -2/3 & 4/3\end{smallmatrix}\right]$. This is exactly the same as the distribution of $\left[\begin{smallmatrix}
\frac{1}{3} & \frac{2}{3} & -\frac{1}{3} \\ \frac{1}{3} & \frac{-1}{3} & \frac{2}{3}\end{smallmatrix}\right]\mech_a(\datavec)$ and so $\mech_b$ is answerable from $\mech_a$. 

Similarly, the output distribution of $\mech_a(\datavec)$ is the same as the distribution of $\left[\begin{smallmatrix}1 & 1\\ 1 & 0\\
0 & 1\end{smallmatrix}\right]\mech_b(\datavec) + 
\sqrt{\frac{2}{3}}\left[\begin{smallmatrix}1 & 0 & 0\\ -1 & 0 & 0\\ -1 & 0 & 0\end{smallmatrix}\right]
N\left(\left[\begin{smallmatrix}
0\\0\\0\end{smallmatrix}\right], \left[\begin{smallmatrix}
1 & 0 & 0\\ 0 & 1 & 0\\ 0 & 0 & 1\end{smallmatrix}\right]\right)$ and thus $\mech_a$ is also answerable from $\mech_b$.
\end{example}

\begin{remark}\label{rem:answer}
It is particularly noteworthy that even though $\mech_a$ and $\mech_b$ in Example \ref{ex:answer} are answerable from each other, $\mech_a$ is obtained from $\mech_b$ by linear postprocessing \underline{followed by noise addition}. The reason is because the rows of the query matrix $\bmat_a=\left[\begin{smallmatrix}1 & 1\\ 1 & 0\\
0 & 1\end{smallmatrix}\right]$ in mechanism $\mech_a$ are \underline{linearly dependent}. The sole purpose of the noise is to convert the 2-dimensional Gaussian distribution obtained by linear postprocessing of $\mech_b$ into the 3-dimensional Gaussian distribution that $\mech_a$ uses. This noise does not add or remove privacy, since $\mech_b$ can also be obtained from $\mech_a$ by linear postprocessing.

Another way to view this phenomenon is to note that the linear dependency in $\bmat_a$ causes inconsistency -- the first component of the output of $\mech_a$ is a noisy sum, and the sum of the 2nd and 3rd components is also a version of the noisy sum. Enforcing consistency \cite{hay2009boosting,LMHMR15} would convert $\mech_a$ into $\mech_b$, and the noise that is removed by consistency is the same noise that is added back when reconstructing $\mech_a$ from $\mech_b$.
\end{remark}

One observation we make from Example \ref{ex:answer} is that proving answerability can be cumbersome because one must produce the matrices $\mat{A}$ and $\mat{C}$ as in Definition \ref{def:answer}. The following result allows us to check answerability in a more mechanical way.

\begin{theoremEnd}[category=section3, all end]{lemma}\label{lemma:aat}
For any real matrix $\mat{A} \in \mathbb{R}^{m \times n}$, if $\mat{A} \mat{A}^T \loewner \mat{I}_m$, then $\mat{A}^T \mat{A} \loewner \mat{I}_n$.
\end{theoremEnd}
\begin{proofEnd}
 Let $r$ be the rank of $\mat{A}$. Using the singular value decomposition, represent $\mat{A}$ as $\mat{A} = \mat{U} \mat{D} \mat{V}$, where $\mat{U}$ is a $\numquery \times m$ orthogonal matrix, $\mat{V}$ is a $n \times n$ orthogonal matrix, and $\mat{D}$ is an $m \times n$ matrix with the singular values $(s_1, s_2, \cdots, s_r)$ of $\mat{A}$ along its main diagonal (the other entries are 0). First, we are given that:
\begin{align*}
    \mat{A} \mat{A}^T & = \mat{U} \mat{D} \mat{D}^T \mat{U}^T \\
    & = \mat{U}~ Diag_{m \times m}(s_1^2, s_2^2, \cdots, s_r^2, 0, \cdots, 0) ~\mat{U}^T \\
    & \loewner~ \mat{I}_m
\end{align*}
Noting that $\mat{U}^T \mat{U} = \mat{I}_m$,
 we have
\begin{align*}
\mat{U}^T\mat{A} \mat{A}^T \mat{U} &=
\mat{U}^T (\mat{U}~ Diag_{m \times m}(s_1^2, s_2^2, \cdots, s_r^2, 0, \cdots, 0) ~\mat{U}^T) \mat{U}\\
&=     Diag_{m \times m}(s_1^2, s_2^2, \cdots, s_r^2, 0, \cdots, 0)\\ 
&\loewner ~\mat{U}^T\mat{I}_m \mat{U} = \mat{I}_m
\end{align*}
which means that $s_i^2 \leq 1$ for $i=1,2, \cdots, r$. Therefore,
\begin{align*}
    \mat{A}^T \mat{A} &= \mat{V}^T \mat{D}^T \mat{D} \mat{V} \\
    & = \mat{V}^T Diag_{n \times n}(s_1^2, s_2^2, \cdots, s_r^2, 0, \cdots, 0) \mat{V} \\
    & \loewner \mat{V}^T \mat{I}_n \mat{V} = \mat{V}^T \mat{V} = \mat{I}_n
\end{align*}
since $\mat{I}_n - Diag_{n \times n}(s_1^2, s_2^2, \cdots, s_r^2, 0, \cdots, 0)$ is a diagonal matrix with nonnegative diagonals, and hence positive semidefinite.
\end{proofEnd}

\begin{theoremEnd}[category=section3, all end]{lemma}\label{lemma:rowspace}
Let $ r(\mat{B})$ denote the row space of a matrix $\mat{B}$. If $\mat{A}$ and  $\mat{B}$ are two symmetric positive semi-definite matrices and $\mat{B} \loewner \mat{A}$, then $r(\mat{B}) \subseteq r(\mat{A})$.
\end{theoremEnd}
\begin{proofEnd}
Consider the linear space of vectors that are orthogonal to all of the rows of $\mat{A}$ and let $\mat{C}$ be a matrix whose columns consist of a full basis  for this space (hence the columns are independent). Then $\mat{A}\mat{C}=\mat{0}$ and a vector $\vec{y}^T$ belongs to the rowspace of $\mat{A}$ if and only if $\vec{y}^T\mat{C}=\mat{0}$.

If $\mat{A}-\mat{B}$ is positive semidefinite, then so is $\mat{C}^T\mat{A}\mat{C} - \mat{C}^T\mat{B}\mat{C}$. But
$\mat{C}^T\mat{A}\mat{C} - \mat{C}^T\mat{B}\mat{C} = -\mat{C}^T\mat{B}\mat{C}$ which is negative semidefinite, and so  $-\mat{C}^T\mat{B}\mat{C}$ is both positive and negative semidefinite, meaning that its eigenvalues are all nonnegative and nonpositive, meaning that they are 0 and $\mat{C}^T\mat{B}\mat{C}=\mat{0}$. 

Since $\mat{B}$  is a symmetric positive semidefinite matrix of rank (say) $r$, it can be written as $\mat{B}=\mat{V}\mat{D}\mat{V}^T$ where $\mat{D}$ is an $r\times r$ diagonal matrix containing the $r$ positive eigenvalues of $\mat{B}$ (if all eigenvalues are $0$ the lemma is trivially true) and $\mat{V}$ is an $m \times r$ matrix whose columns are orthogonal.

Thus if $\mat{C}^T\mat{B}\mat{C}=\mat{0}$ then $\mat{C}^T\mat{V}\mat{D}\mat{V}^T\mat{C}=\mat{0}$ and so for any column vector $\vec{c}_i$ of $\mat{C}$, 
\begin{align*}
0 &=\vec{c}_i^T \mat{V}\mat{D}\mat{V}^T\vec{c}_i\\
  &= \sum_j (\vec{c}_i \cdot \vec{v}_j)^2 \mat{D}[j,j] \quad\text{where $\vec{v}_j$ is the $j^\text{th}$ column of $\mat{V}$}
\end{align*}
Noting that the diagonals of $\mat{D}$ are all positive, this means that $\vec{v}_j^T\vec{c}_i=0$ for all $i$ and $j$. Thus $\mat{B}\mat{C} = \mat{V}\mat{D}\mat{V}^T\mat{C}=\mat{0}$ and therefore the rows of $\mat{B}$ belong to the row space of $\mat{A}$.
\end{proofEnd}

\begin{theoremEnd}[category=section3, all end]{lemma}\label{lemma:subset}
Let $ r(\mat{B})$ denote the row space of the matrix $\mat{B}$. If $r(\mat{B}) \subseteq r(\mat{A})$, then $\mat{B} = \mat{B} \mat{A}^+ \mat{A}$, where $\mat{A}^+$ is the Moore-Penrose pseudoinverse of $\mat{A}$.
\end{theoremEnd}
\begin{proofEnd}
If $r(\mat{B}) \subset r(\mat{A})$ then $\mat{B} =\mat{C}\mat{A}$ for some matrix $\mat{C}$. Then 
$$\mat{B}\mat{A}^+\mat{A} = \mat{C}\mat{A}\mat{A}^+\mat{A}=\mat{C}\mat{A}=\mat{B}$$
by the property of pseudoinverses ($\mat{A}\mat{A}^+\mat{A}=\mat{A}$).
\end{proofEnd}

\begin{theoremEnd}[category=section3]{theorem}
\label{thm:answer}
Lets $\mech_a(\datavec) = \bmat_a \datavec + N(\vec{0}, \covar_a)$ and $\mech_b(\datavec) = \bmat_b \datavec + N(\vec{0}, \covar_b)$ be linear Gaussian mechanisms.
 $\mech_b$ is answerable from $\mech_a$ if and only if $\bmat_b^T\covar_b^{-1} \bmat_b ~\loewner~  \bmat_a^T \covar_a^{-1} \bmat_a $ (i.e., $\bmat_a^T \covar_a^{-1} \bmat_a - \bmat_b^T\covar_b^{-1} \bmat_b$ is positive semidefinite, and hence its eigenvalues are nonnegative).
\end{theoremEnd}
\begin{proofEnd}
First, we prove that ``if'' direction. Suppose $\mech_b$  is answerable from $\mech_a$. 
By definition, there exist  matrices $\mat{A}$ and $\mat{C}$ such that
\begin{align*}
    \bmat_b & = \mat{A} \bmat_a \\
    \covar_b &= \mat{A} \covar_a \mat{A}^T + \mat{C}\mat{C}^T\\
    &\text{ and so}\\
    \mat{A} \covar_a \mat{A}^T & \loewner~\covar_b 
\end{align*}

Let $\bmat_{a \star} = \covar_a^{-1/2} \bmat_a$ (where $\covar_a^{1/2}$ is the inverse of a symmetric positive definite matrix square root of $\covar_a$) and $\bmat_{b \star} = \covar_b^{-1/2} \bmat_b$. Then let $\mat{A}_{\star} = \covar_b^{-1/2} \mat{A} \covar_a^{1/2}$, then we have 
\begin{align*}
    \bmat_{b \star} &= \covar_b^{-1/2} \bmat_b\\
    &= \covar_b^{-1/2} \mat{A} \bmat_a\\
    &= \covar_b^{-1/2} \mat{A} \covar_a^{1/2}\covar_a^{-1/2}\bmat_a\\
    & = \mat{A}_{\star} \bmat_{a \star} \\
     \mat{A}_{\star} \mat{A}_{\star}^T &= \left(\covar_b^{-1/2} \mat{A} \covar_a^{1/2}\right)\left(\covar_b^{-1/2} \mat{A} \covar_a^{1/2}\right)^T\\ 
     &= \covar_b^{-1/2} \mat{A} \covar_a \mat{A}^T \covar_b^{-1/2}  \\
     &\loewner~  \covar_b^{-1/2}  \covar_b  \covar_b^{-1/2}\quad(\text{ since } \mat{A} \covar_a \mat{A}^T  \loewner~\covar_b)\\
    &= \mat{I}_{m_b} \quad\text{(where $m_b$ is the number of rows of $\bmat_b$)} 
\end{align*}
This means that $ \mat{A}_{\star} \mat{A}_{\star}^T \loewner \mat{I}_{m_b}$ and so by
Lemma \ref{lemma:aat} we know that $\mat{A}_{\star}^T \mat{A}_{\star} \loewner \mat{I}_{m_a}$, where $m_a$ is the number of rows of $\bmat_a$. Thus
\begin{align*}
    \bmat_b^T \covar_b^{-1} \bmat_b &= \bmat_a^T\mat{A}^T \covar_b^{-1} \mat{A}\bmat_a\\
    &=(\bmat_a^T\covar_a^{-1/2})\covar_a^{1/2}\mat{A}^T \covar_b^{-1} \mat{A}\covar_a^{1/2}(\covar_a^{-1/2}\bmat_a)\\
    &=(\bmat_a^T\covar_a^{-1/2})(\covar_a^{1/2}\mat{A}^T \covar_b^{-1/2})(\covar_b^{-1/2} \mat{A}\covar_a^{1/2})(\covar_a^{-1/2}\bmat_a)\\
    & =   \bmat_{a \star}^T \mat{A}_{\star}^T \mat{A}_{\star} \bmat_{a \star} \\
    & \loewner ~\bmat_{a \star}^T \mat{I}_{m_a} \bmat_{a \star}\quad(\text{since }\mat{A}_{\star}^T \mat{A}_{\star} \loewner \mat{I}_{m_a}) \\ 
    & = \bmat_{a \star}^T  \bmat_{a \star} \\
    &= \bmat_a^T \covar_a^{-1} \bmat_a
\end{align*}
so if $\mech_b$ is answerable from $\mech_a$, we have $\bmat_b^T \covar_b^{-1} \bmat_b \loewner \bmat_a^T \covar_a^{-1} \bmat_a$

Now we must prove the other direction (i.e., 
if $\bmat_b^T \covar_b^{-1} \bmat_b \loewner  \bmat_a^T \covar_a^{-1} \bmat_a $, we must show  $\mech_b$ is answerable from $\mech_a$).

As before, let $\bmat_{a \star} = \covar_a^{-1/2} \bmat_a$ (where $\covar_a^{1/2}$ is the inverse of a symmetric positive definite matrix square root of $\covar_a$) and $\bmat_{b \star} = \covar_b^{-1/2} \bmat_b$. Consider the mechanisms:
\begin{align*}
    \mech_{a \star} (x) &= \bmat_{a \star} + N(0, \mat{I}_{m_a}) \\
    \mech_{b \star} (x) &= \bmat_{b \star} + N(0, \mat{I}_{m_b})
\end{align*}
We have
\begin{align*}
    \bmat_{b \star}^T \bmat_{b \star}  &= \bmat_b^T \covar_b^{-1} \bmat_b \\
    & \loewner  \bmat_a^T \covar_a^{-1} \bmat_a \\
    & = \bmat_{a \star}^T \bmat_{a \star}
\end{align*}

Let $r(\mat{C})$ denote the row space of a matrix $\mat{C}$. From Lemma \ref{lemma:rowspace} we know that 
\begin{align*}
    r(\bmat_{b \star})  = r(\bmat_{b \star}^T \bmat_{b \star}) \subseteq r(\bmat_{a \star}^T \bmat_{a \star})  =r(\bmat_{a \star})  
\end{align*}

Let $\mat{A}_{\star} = \bmat_{b \star} \bmat_{a \star}^{+}$, then from lemma \ref{lemma:subset} we have $\bmat_{ b \star} = \mat{A}_{\star} \bmat_{a \star}$.

Now, the dimensions of $\bmat_a$ and $\bmat_b$ (and hence $\bmat_{a*}$ and $\bmat_{b*}$ are $\numquery_a \times \dimsize$ and $\numquery_b\times \dimsize$ (they have the same number of columns because they are defined for mechanisms whose input is a vector $\datavec$ with $\dimsize$ components).

Taking the SVD, we can get the following representation:
\begin{align*}
    \bmat_{b \star} & = \mat{U}_b \mat{D}_b \mat{V}_b  \\
    \bmat_{a \star} & = \mat{U}_a \mat{D}_a \mat{V}_a \\
    \intertext{ and so}
    \bmat_{a \star}^+ & = \mat{V}_a^T \mat{D}_a^+ \mat{U}_a^T 
\end{align*} 
where $\mat{U}_a$ (resp., $\mat{U}_b$)is an $\numquery_a\times \numquery_a$ (resp., $\numquery_b\times \numquery_b$) orthogonal matrix, $\mat{D_a}$ (resp., $\mat{D}_b$) is an $\numquery_a\times\dimsize$ (resp., $\numquery_b\times\dimsize$) matrix, and both $\mat{V}_a$ and $\mat{V}_b$ are $\dimsize \times\dimsize$ orthogonal matrices.

Since $\bmat_{b \star}^T \bmat_{b \star}\loewner \bmat_{a \star}^T \bmat_{a \star} $ we know that
\begin{align*}
    \mat{V}_b^T \mat{D}_b^T \mat{D}_b \mat{V}_b \loewner \mat{V}_a^T \mat{D}_a^T \mat{D}_a \mat{V}_a
\end{align*}
and 
\begin{align*}
    \mat{A}_{\star}^T \mat{A}_{\star} & = \left(\bmat_{a \star}^{+ T} \bmat_{b \star}^T\right) \left( \bmat_{b \star} \bmat_{a \star}^+ \right)\\
    & = \left( \mat{U}_a \mat{D}_a^{+ T} \mat{V}_a \mat{V}_b^T \mat{D}_b^T \mat{U}_b^T \right) \left( \mat{U}_b \mat{D}_b \mat{V}_b \mat{V}_a^T \mat{D}_a^+ \mat{U}_a^T \right)\\
    & = \mat{U}_a \mat{D}_a^{+ T} \mat{V}_a \underline{\mat{V}_b^T \mat{D}_b^T \mat{D}_b \mat{V}_b} \mat{V}_a^T \mat{D}_a^+ \mat{U}_a^T \\
    & \loewner ~ \mat{U}_a \mat{D}_a^{+ T} \mat{V}_a \underline{\mat{V}_a^T \mat{D}_a^T \mat{D}_a \mat{V}_a} \mat{V}_a^T \mat{D}_a^+ \mat{U}_a^T \\
    & = \mat{U}_a \mat{D}_a^{+ T}  \mat{D}_a^T \mat{D}_a \mat{D}_a^+ \mat{U}_b^T 
\end{align*}
Because $\mat{D}_a$ is a $\numquery_a \times \dimsize$ diagonal matrix, where the elements of its main diagonal are the singular values $s_i$ of $\bmat_a$, then $\mat{D}_a^+$ is a $\dimsize \times \numquery_a$ diagonal matrix, where the elements of its main diagonal are $\frac{1}{s_i}$ for the nonzero singular values (and $0$ otherwise). So, that $\mat{D}_a \mat{D}_a^+$ is a $\numquery_a \times \numquery_a$ diagonal matrix where the elements of its diagonal are either 1 or 0. The same thing holds for $\mat{D}_a^{+T} \mat{D}_a^T$. Therefore, $\mat{D}_a \mat{D}_a^+ \mat{D}_a^{+T} \mat{D}_a^T \loewner \mat{I}_{\numquery_a}$ and 
\begin{align*}
    \mat{A}_{\star}^T \mat{A}_{\star} &\loewner \mat{U}_a \mat{I}_{\numquery_a} \mat{U}_a^T \\
    & = \mat{U}_a \mat{U}_a^T = \mat{I}_{\numquery_a}
\end{align*}
From Lemma \ref{lemma:aat} we have
\begin{align*}
    \mat{A}_{\star} \mat{A}_{\star}^T \loewner \mat{I}_{\numquery_b}
\end{align*}
Let $\mat{A} = \covar_b^{1/2} \mat{A}_{\star} \covar_a^{-1/2}$, we can see that 
\begin{align*}
    \bmat_b & = \covar_b^{1/2} \bmat_{b \star}\\ & = \covar_b^{1/2} \mat{A}_{\star} \bmat_{a \star}\quad\text{which we showed previously}\\
    & = \covar_b^{1/2}  \mat{A}_{\star} \covar_a^{-1/2} \bmat_{a} \\
    & =\mat{A} \bmat_a \\
    \covar_b & = \covar_b^{1/2} \mat{I}_{\numquery_b} \covar_b^{1/2}  \\
    & \rloewner \covar_b^{1/2} \mat{A}_{\star} \mat{A}_{\star}^T \covar_b^{1/2} \quad\text{since we showed} \mat{A}_{\star} \mat{A}_{\star}^T \loewner \mat{I}_{\numquery_b}\\
    & = \mat{A} \covar_a \mat{A}^T
\end{align*}
Then, for the purposes of answerability in Definition \ref{def:answer}, we have defined the matrix $\mat{A}$. The matrix $\mat{C}$ can be obtained by noting that $\bmat_b - \mat{A}\covar_a\mat{A}^T$ is positive semidefinite and so has a symmetric matrix square root. Setting $\mat{C}$ to be this square root, we have  $\bmat_{b}=\mat{A}\covar_a\mat{A}^T + \mat{C}\mat{C}^T$ and therefore $\mech_b$ is answerable from $\mech_a$ by Lemma \ref{lem:answer}.
\end{proofEnd}

One interesting consequence of Theorem \ref{thm:answer} is that answerability depends on the quantities $\bmat^T_a \covar_a^{-1}\bmat_a$ and $\bmat^T_b \covar_b^{-1}\bmat_b$, which are the privacy cost matrices of mechanisms $\mech_a$ and $\mech_b$, respectively (see Section \ref{subsec:gaussmech}).

In particular, if $\bmat^T_a \covar_a^{-1}\bmat_a=\bmat^T_b \covar_b^{-1}\bmat_b$ then $\mech_a$ and $\mech_b$ are not only answerable from each other, but also have the exact same privacy cost under zCDP, approximate differential privacy, Gaussian differential privacy, and their personalized versions (as well as any other postprocessing invariant privacy definitions). In this sense, they are completely identical in terms of information content and privacy. This leads to the following definition.

\begin{definition}[Equivalence]\label{def:equal}
Two mechanisms $\mech_a$ and $\mech_b$ are \emph{equivalent} if $\mech_a$ is answerable from $\mech_b$ and vice versa.
In particular, if $\mech_a(\datavec)=\bmat_a \datavec + N(\vec{0}, \covar_a)$ and $\mech_b(\datavec)=\bmat_b\datavec + N(\vec{0}, \covar_b)$ are linear Gaussian mechanisms, then $\mech_a$ and $\mech_b$ are \emph{equivalent} if $\bmat^T_a \covar_a^{-1}\bmat_a=\bmat^T_b \covar_b^{-1}\bmat_b$.
\end{definition}

\subsection{Shared Information and Common Mechanisms}\label{subsec:shared}
Intuitively, a piece of information is shared by $\mech_1$ and $\mech_2$ if that information can be derived from the output of $\mech_1$ \emph{and} it can also be derived from the output of $\mech_2$.

We formalize ``information'' as a mechanism $\mech_c$ that can be answered from $\mech_1$ and from $\mech_2$. Such a mechanism is called a \emph{common} mechanism of $\mech_1$ and $\mech_2$.

\begin{definition}[Common Mechanism]\label{def:common}
A mechanism $\mech_c$ is common to $\mech_1$ and $\mech_2$ if $\mech_c$ is answerable from $\mech_1$ and is answerable from $\mech_2$. When they are all linear Gaussian mechanisms: 
\begin{align*}
\mech_1(\datavec) &= \bmat_1 \datavec + N(\vec{0}, \covar_1)\\ \mech_2(\datavec) &= \bmat_2 \datavec + N(\vec{0}, \covar_2) \\ \mech_c(\datavec) &= \bmat_c \datavec + N(\vec{0}, \covar_c)
\end{align*}
then by Theorem \ref{thm:answer}$, \mech_c$ is common to $\mech_1$ and $\mech_2$ whenever:
\begin{align*}
\bmat^T_c \covar_c^{-1}\bmat_c &\loewner \bmat^T_a \covar_a^{-1}\bmat_a \text{ and }\\
\bmat^T_c \covar_c^{-1}\bmat_c &\loewner\bmat^T_b \covar_b^{-1}\bmat_b
\end{align*}
\end{definition}

\begin{example}\label{ex:common}
Consider the following four mechanisms, where $\mech_1,\mech_3,\mech_4$ are noisy sum queries with variances 1, 2, and 1.5, respectively, while $\mech_2$ is a combination of a noisy sum query with variance 2 and a noisy identity query with variance 2:
\begin{align*}
\mech_1(\datavec) &= [\begin{smallmatrix} 1&1&1\end{smallmatrix}]\datavec + N(0, \sigma^2=1)\\
\mech_2(\datavec) &= \left[\begin{smallmatrix} 
                    1&1&1\\
                    1&0&0\\
                    0&1&0\\                       
                    0&0&1
                    \end{smallmatrix}\right]\datavec +
                    N\left(\left[\begin{smallmatrix}
                           0\\0\\0\\0\end{smallmatrix}\right],
                           \left[\begin{smallmatrix}
                              2 & 0 & 0 & 0\\ 0 & 2 & 0 & 0\\ 0 & 0 &
                              2 & 0\\0&0&0&2\end{smallmatrix}\right]\right)\\
\mech_3(\datavec) &= [\begin{smallmatrix} 1&1&1\end{smallmatrix}]\datavec + N(0, \sigma^2=2)\\
\mech_4(\datavec) &= [\begin{smallmatrix} 1&1&1\end{smallmatrix}]\datavec + N(0, \sigma^2=1.5)\\
\end{align*}
The mechanism $\mech_3$ is common to both $\mech_1$ and $\mech_2$ because it can be answered using only the output of either mechanism (with no additional access to the underlying data). For example, $\mech_3$ can be answered by adding noise to the output of $\mech_1$ as follows: $\mech_3(\datavec)=\mech_1(\datavec)+N(0,1)$. Also $\mech_3$ can be answered from $\mech_2$ by taking the noisy sum that $\mech_2$ directly provides (mathematically, $\mech_3(\datavec)=[\begin{smallmatrix} 1&0&0&0\end{smallmatrix}]\mech_2(\datavec)$). However, $\mech_4$ is also common to $\mech_1$ and $\mech_2$ as we can see from the following equations:
\begin{align*}
\mech_4(\datavec) &= \mech_1(\datavec) + N(0, 0.5)\\
\mech_4(\datavec) &= [0.75, 0.25, 0.25, 0.25]\mech_2(\datavec) 
\end{align*}
\end{example}

As we see from Example \ref{ex:common}, both $\mech_3$ and $\mech_4$ are common mechanisms for $\mech_1$ and $\mech_2$, and so they both capture information that is shared by $\mech_1$ and $\mech_2$. However, $\mech_4$ clearly captures more of this shared information than $\mech_3$. This leads to a concept of a \emph{maximally} common mechanism.

\begin{definition}[Maximally common mechanism]\label{def:comech}
A mechanism $\commech$ is maximally common to $\mech_1$ and $\mech_2$ if (1) $\commech$ is common to $\mech_1$ and $\mech_2$, (2) if there is another mechanism $\mech^\dagger$ that is common to $\mech_1$ and $\mech_2$ and if $\commech$ is answerable from $\mech^\dagger$, then $\commech$ and $\mech^\dagger$ are equivalent.
\end{definition}

It turns out that $\mech_4$ is a maximally common mechanism to $\mech_1$ and $\mech_2$. We show how to compute maximally common mechanisms in Section \ref{sec:common}.

\subsection{Decomposition into a Common and Residual Mechanism}\label{subsec:decompose}
Now that a maximal common mechanism $\commech$ for $\mech_1$ and $\mech_2$ has been defined, we next define the residual mechanisms $\mech_1^\prime$ and $\mech_2^\prime$. Intuitively, $\mech_1^\prime$ (resp., $\mech_2^\prime$) represents the least amount of additional information that, when combined with $\commech$ allows us to recreate $\mech_1$ (resp., $\mech_2$). Alternatively, $\mech^\prime_1$ (resp., $\mech_2^\prime$) is the result of ``subtracting away'' the information about $\commech$ from $\mech_1$ (resp., $\mech_2$). Recalling that the notation $(\commech, \mech_1^\prime)$ is a mechanism that runs both $\commech$ and $\mech_1^\prime$ on the data and releases their outputs, we  can formally define residual mechanisms as follows:

\begin{definition}[Residual Mechanism]\label{def:residual}
Given mechanisms $\mech_1$, $\mech_2$ and a maximally common mechanism $\commech$, we say that $\mech_1^\prime$ and $\mech_2^\prime$ are \emph{residual} mechanisms if:
\begin{itemize}
\item $(\commech, \mech_1^\prime)$ is equivalent (see Definition \ref{def:equal}) to $\mech_1$ and
\item $(\commech, \mech_2^\prime)$ is equivalent  to $\mech_2$.
\end{itemize}
\end{definition}

Note that, by virtue of equivalence, $(\commech, \mech_1^\prime)$ has the same privacy cost as $\mech_1$ under any postprocessing-invariant privacy definition (including all the ones studied in this paper), and similarly with $(\commech, \mech_2^\prime)$ and $\mech_2$.
The checking of equivalence between $(\mech_*, \mech_1^\prime)$ and $\mech_1$ can be done using the following result:

\begin{theoremEnd}[category=section3]{lemma}\label{lem:residual}
Suppose that  $\mech_1(\datavec)=\bmat_1 \datavec + N(\vec{0}, \covar_1)$ and  $\mech_2(\datavec)=\bmat_2\datavec + N(\vec{0}, \covar_2)$ are linear Gaussian mechanisms and that $\commech(\datavec)=\comb \datavec + N(\vec{0}, \comvar)$ is their maximally common mechanism. Then $\mech_1^\prime(\datavec)=\bmat_1^\prime \datavec + N(\vec{0}, \covar_1^\prime)$ and $\mech_2^\prime(\datavec)=\bmat_2^\prime\datavec + N(\vec{0}, \covar_2^\prime)$ are residual mechanisms if and only if:
\begin{align*}
\comb^T \comvar^{-1}\comb ~+~ (\bmat_1^\prime)^T (\covar_1^\prime)^{-1} \bmat_1^\prime ~&= ~\bmat_1^T \covar_1^{-1} \bmat_1\\
\comb^T \comvar^{-1}\comb ~+~ (\bmat_2^\prime)^T (\covar_2^\prime)^{-1} \bmat_2^\prime ~&= ~\bmat_2^T \covar_2^{-1} \bmat_2\\
\end{align*}
In which case $(\commech, \mech_1^\prime)$ is equivalent to $\mech_1$ and $(\commech, \mech_2^\prime)$ is equivalent to $\mech_2$.
\end{theoremEnd}
\begin{proofEnd}
We prove the result for $\mech_2$ as the result for $\mech_2$ is similar. The combined mechanism $(\commech, \mech_1^\prime)$ can be expressed in matrix form as follows:
\begin{align*}
\left[
\begin{smallmatrix}
\comb\\\bmat_1^\prime
\end{smallmatrix} 
\right]\datavec + N\left(\vec{0}, 
\left[\begin{smallmatrix}
\comvar &\mat{0}\\
\mat{0} & \covar_1^\prime
\end{smallmatrix}\right]\right)
\end{align*}
Thus $(\commech, \mech_1^\prime)$ is a linear Gaussian mechanism with $\bmat=\left[
\begin{smallmatrix}
\comb\\\bmat_1^\prime
\end{smallmatrix} 
\right]$
and $\covar = \left[\begin{smallmatrix}
\comvar &\mat{0}\\
\mat{0} & \covar_1^\prime
\end{smallmatrix}\right]$.
The privacy cost is
\begin{align*}
\bmat^T \covar^{-1}\bmat &= \comb^T \comvar^{-1}\comb ~+~ (\bmat_1^\prime)^T (\covar_1^\prime)^{-1} \bmat_1^\prime 
\end{align*}
and the rest follows directly from Theorem \ref{thm:answer}.
\end{proofEnd}

\begin{example}
Consider a table with 2 attributes $Att_1$ and $Att_2$, each attribute taking 3 possible values $a,b,c$. The data vector $\datavec$ then has 9 components. A marginal on $Att_1$ then consists of 3 numbers: the number of people for which $Att_1=a$, the number of people for which $Att_1=b$, and the number of people for which $Att_1=c$. The marginal on $Att_2$ is defined analogously. Consider a mechanism $\mech_1$ that adds independent $N(0,1)$ noise to the marginal on $Att_1$ and a mechanism $\mech_2$ that adds independent $N(0,1)$ noise to the marginal on $Att_2$. In matrix notation, they are represented as follows:
\begin{align*}
\mech_1(\datavec) &= 
\left[\begin{smallmatrix}
1&1&1 & 0&0&0 & 0&0&0\\
0&0&0 & 1&1&1 & 0&0&0\\
0&0&0 & 0&0&0 & 1&1&1\\
\end{smallmatrix}\right]\datavec + 
N\left(
\left[\begin{smallmatrix}0\\0\\0\end{smallmatrix}\right],
\left[\begin{smallmatrix}1 & 0 & 0\\0 & 1& 0\\0&0&1\end{smallmatrix}\right]
\right)\\
\mech_2(\datavec) &= 
\left[\begin{smallmatrix}
1&0&0 & 1&0&0 & 1&0&0\\
0&1&0 & 0&1&0 & 0&1&0\\
0&0&1 & 0&0&1 & 0&0&1\\
\end{smallmatrix}\right]\datavec + 
N\left(
\left[\begin{smallmatrix}0\\0\\0\end{smallmatrix}\right],
\left[\begin{smallmatrix}1 & 0&0\\0 & 1&0\\0&0&1\end{smallmatrix}\right]
\right)
\end{align*}
From the output of $\mech_1$ we can add up the noisy counts of people having values $a$, $b$, and $c$ for attribute $Att_1$ to get an estimate of the total number of people. This estimate has variance 3. We can do the same with $\mech_2$ to get a noisy total with variance 3. Running either mechanism thus provides  a noisy total with variance 3, and the noisy total mechanism is in fact their maximal common mechanism, and is represented as:
$$\commech(\datavec) =\left[\begin{smallmatrix}
1&1&1 & 1&1&1 & 1&1&1\\
\end{smallmatrix}\right]\datavec + N(0, 3).$$
The corresponding residual mechanisms are:
\begin{align*}
\mech_1^\prime(\datavec) =\left[\begin{smallmatrix}
 0  & 0 &  0 & 1 & 1&  1 & -1 & -1 & -1\\
 -1 & -1 & -1 & 1 & 1 & 1 &  0 &  0 &  0
\end{smallmatrix}\right]\datavec + N(
\left[\begin{smallmatrix}0\\0\end{smallmatrix}\right],
\left[\begin{smallmatrix}2 & 1\\  1 & 2
\end{smallmatrix}\right])\\
\mech_2^\prime(\datavec) =\left[\begin{smallmatrix}
   0 & 1  &-1 &  0 & 1 & -1 &  0 & 1 & -1\\
 -1 & 1 &  0 & -1 & 1 &  0 & -1 & 1 &  0
\end{smallmatrix}\right]\datavec + N(
\left[\begin{smallmatrix}0\\0\end{smallmatrix}\right],
\left[\begin{smallmatrix}2 & 1\\  1 & 2
\end{smallmatrix}\right])
\end{align*}
The residual mechanism $\mech_1^\prime$ is answering two queries: (1) \# of records with $Att_1=b$ minus the \# with $Att_1=c$, and (2) \# of records with $Att_1=b$ minus the \# with $Att_1=a$. Both queries get variance 2 and are correlated with covariance 1. The residual mechanism $\mech_2^\prime$ works analogously for attribute $Att_2$.

The original mechanism $\mech_1$ can be recovered from the outputs of  $\mech_1^\prime$ and $\commech$ as follows:
$$\mech_1(\datavec) = 
\left[\begin{smallmatrix}1/3 & -2/3\\1/3 & 1/3\\-2/3 & 1/3\end{smallmatrix}\right]
\mech_1^\prime(\datavec) +
\left[\begin{smallmatrix}1/3\\1/3\\1/3\end{smallmatrix}\right]
\commech(\datavec)$$
and it is easy to check using Lemma \ref{lem:residual} that $\mech_1$ and $(\commech, \mech_1^\prime)$ are indeed equivalent (providing the same information and having the same privacy cost).
\end{example}
We show how to compute the residual mechanisms in Section \ref{sec:common}.

\subsection{Formalizing Decision Making with the Common Mechanism}\label{subsec:statement}

Having formally defined the, (maximally) common mechanism $\commech$ of linear Gaussian mechanisms $\mech_1$ and $\mech_2$, and having defined the residual mechanisms $\mech_1^\prime$ and $\mech_2^\prime$, the formal problem statement can be defined as the following sequence of steps:
\begin{enumerate}[leftmargin=*]
\item Given two linear Gaussian query mechanisms $\mech_1$ and $\mech_2$, compute their maximal common mechanism $\commech$ (algorithms are provided in Section \ref{sec:common}).
\item Given $\mech_1,\mech_2$ and $\commech$, compute the residual mechanisms $\mech_1^\prime$ and $\mech_2^\prime$ (algorithms are provided in Section \ref{sec:common}).
\item Run $\commech$ on the data to produce an output $\outp_*$.
\item Based on $\outp_*$, decide whether to run the residual mechanism $\mech_1^\prime$ or $\mech_2^\prime$. An analyst is free to choose how to make a decision based on $\outp^*$, but for more automated approaches, we provide suggestions in Section \ref{sec:decide}.
\item Based on the decision, either run $\mech_1^\prime$ on the data (and combine the result with $\outp_*$ to obtain an answer to $\mech_1$) or run $\mech_2^\prime$ on the data (and combine the result with $\outp_*$ to obtain an answer to $\mech_2$). Algorithms for recovering  $\mech_1$ from $(\commech, \mech_1^\prime)$ and recovering $\mech_2$ from $(\commech, \mech_2^\prime)$ are given in Section \ref{sec:common}.
\end{enumerate}

\subsection{Additional Potential Uses}\label{subsec:additional}
We believe that the idea of explicitly factoring out information that is common to a set of mechanisms can be used widely as another tool in differential privacy applications. While we describe one application in this paper, there are additional possibilities for future work that we describe here.

\subsubsection{Reducing privacy cost for systems that interact with analysts.} 
When a data analyst queries a differentially private system in a sequential manner, the system may wish to use previously cached noisy answers to help answer new queries, thus reducing the privacy cost of answering the new queries. This is one of the goals of systems such as Apex \cite{apex,cacheme}. The use of common mechanisms could help such systems achieve their goals by factoring out what is common between the new queries and previously asked queries, computing the residual mechanism for the new queries. The residual mechanism captures the information in the new queries that are not contained in the previous queries. The system would then only need to expend the privacy budget on the residual mechanism instead of the entire set of new queries.

\subsubsection{Enhancing the exponential mechanism.}
When one query needs to be privately selected from a set $\{\query_1,\dots,\query_k\}$ of $k$ queries, the exponential mechanism \cite{exponentialMechanism} can often be used for this purpose. When the queries have common substructures, part of the privacy budget for the exponential mechanism can be used to select among the substructures, answer them, and then use this information and the remaining privacy budget to design an exponential mechanism quality function for choosing among the residual mechanisms. This possibility is an area of future work.

\subsubsection{Synthetic data generation.}
There are several ways that the common mechanism can be used to enhance iterative algorithms for generating synthetic data (e.g., \cite{hardt2012simple,liu2021iterative,aydore2021differentially,mckenna2022aim}) that first privately select queries and then use their noisy answers to update a data synopsis. First, one can take mechanisms that are common to subsets of the original queries, and add them to the list of queries that can be selected in each round. Another use is to take the previously selected queries from earlier rounds and to ``remove'' their information from the remaining queries (via residual mechanisms). For example, if a one-way marginal Age had been answered in an earlier round, any information answerable from this marginal could be removed from the remaining queries (i.e., replaced by query matrices of residual mechanisms), so that the remaining queries focus on information that has not yet been provided. Thus integrating different uses of the common mechanism framework into synthetic data generation is another direction for future work.

\section{Related Work}\label{sec:related}
There are roughly two types of tasks that differentially private algorithms perform: (1) query \emph{selection}, which involves deciding \emph{which} queries need answers, and (2) query \emph{measurement} in which noisy answers to the queries are created.

Query selection is an important problem, with many important applications, such as synthetic data generation \cite{hardt2012simple,mckenna2022aim,cai2021data,privbayes,aydore2021differentially,liu2022towards,liu2021iterative,mckenna2021winning,liu2021leveraging},  as well as hyperparameter tuning \cite{diffperm,LiuPSPC},  feature selection \cite{GuhaSmith13}, frequent itemset mining \cite{BhaskarDFP}, exploring a privacy/accuracy tradeoff \cite{LigettNRWW17}, data pre-processing \cite{ChenDPRD}, PAC learning \cite{bun2019private}, etc.

Query selection can be performed in a \emph{data-independent} way, meaning that the queries are chosen in advance and no privacy loss budget is spent on the choice. In the case of linear queries, the techniques that plan the optimal set of queries in advance are generally called \emph{matrix mechanisms} \cite{LMHMR15,mckenna2018optimizing,YYZH16,YZWXYH12,yuan2015optimizing,mckenna2018optimizing,xiao2020optimizing}. Matrix mechanisms \emph{implicitly} take advantage of shared information between queries by setting up an optimization problem that finds a query strategy matrix that can answer the prespecified queries as accurately as possible under a privacy constraint.

Query selection can also be performed in a data-dependent way by allocating some of the privacy loss budget to specially designed selection mechanisms. The sparse vector technique \cite{diffpbook,lyu2017understanding}, exponential mechanism \cite{exponentialMechanism} and  various generalizations \cite{steinke2017tight,LigettNRWW17,BeimelNS16,GuhaSmith13,RaskhodnikovaS16,ChaudhuriLMM,LiuPSPC,freegap,freegap2} are commonly used for this task, along with bespoke algorithms targeted at specific applications \cite{ahp,php,sf,li2014data}.

Head-to-head comparisons between data-dependent and data-independent methods show that there is no clear winner \cite{dpbench} -- in some situations, data-dependent selection provides the best choice of queries. In other situations, data-independent methods prevail. Thus it is important to keep expanding the available toolkit for query selection.

There is also a much smaller class of zero-waste differentially private selection algorithms \cite{xiao2011ireduct,koufogiannis2015gradual,li2011enabling,xiao09opt} whose purpose is to adaptively determine how much noise to add to a query, without wasting privacy loss budget. For example, suppose one is interested in the count of the number of people over 18. An analyst is prepared to spend up to $\epsilon_1$ of her privacy loss budget (using pure $\epsilon$-differential privacy) to get the answer. But, if the true number is large, she would prefer to use a smaller privacy loss budget value $\epsilon_2$ (this results in a higher absolute error but is tolerable when the true answer is large because it still results in a small relative error). NoiseDown, which was introduced by Xiao et al. \cite{xiao2011ireduct} (but had a bug in the algorithm) and later corrected by Koufogiannis et al. \cite{koufogiannis2015gradual}, is one technique to solve this problem. The analyst adds Laplace noise with privacy budget $\epsilon_2$ to the true answer. Based on this answer, the analyst can either keep it or refine the noise so that the total privacy loss budget is $\epsilon_1$ and so that the accuracy is the same as if she had used Laplace noise with budget $\epsilon_1$ in the first place. Variations for this single-query noise refinement were also studied for randomized response \cite{xiao09opt} and Gaussian noise \cite{li2011enabling}.

Our approach is also zero-waste, while being much more general. Instead of choosing between two versions of the same mechanism (the only difference being its privacy cost/amount of noise), our method allows the choice between two or more linear mechanisms that use Gaussian noise (and the mechanisms may also have different privacy costs). This adds another tool to the algorithmic toolbox for differential privacy. Although not necessarily a replacement for the exponential mechanism, we show empirically that our approach is useful in situations where it is difficult to specify the quality function that the exponential mechanism needs. One direction of future work is to combine the exponential mechanism with our technique -- using the output of the common mechanism to fine-tune the construction of the exponential mechanism.

\section{Algorithms} \label{sec:common}
We next show how to compute the common mechanism (Section  \ref{subsec:makecommon}), compute the residual mechanisms (Section \ref{subsec:makeresidual}), and recreate the original mechanism from the common and residual mechanism (Section \ref{subsec:recreate}).

\subsection{Computing the Common Mechanism} \label{subsec:makecommon}
The full procedure for computing the common mechanism is shown in Algorithm \ref{alg:commech}. We now explain how it is derived.

\begin{algorithm}
   \DontPrintSemicolon
    \KwIn{Linear Gaussian Mechanisms $\mech_{1} $, $\mech_{2}$}
    $\bmat_{1}\gets$  Standardization($\mech_1$) \tcp*{see Algorithm \ref{alg:standardize}}\label{line:standard1}
    $\bmat_{2}\gets$  Standardization($\mech_2$)\tcp*{see Algorithm \ref{alg:standardize}}\label{line:standard2}
    $\comb \gets $ basis for $\rowspace(\bmat_1)\cap\rowspace(\bmat_2)$\;\label{line:basis}
    \tcp{Calculate the covariance matrix \covar}
    $\mat{A}_{1}  \gets \bmat \bmat_{1}^{\dagger} $\tcp*{$\dagger$ is the Moore-Penrose pseudoinverse}
    $\mat{A}_{2}  \gets \bmat \bmat_{2}^{\dagger} $\;
    $\comvar \gets \frac{\mat{A}_{1}  \mat{A}_{1}^T + \mat{A}_{2}  \mat{A}_{2}^T}{2} + \frac{|\mat{A}_{2}  \mat{A}_{2}^T - \mat{A}_{1}  \mat{A}_{1}^T|}{2}$\tcp*{where $|\cdot|$ replaces negative eigenvalues in a matrix with positive eigenvalues}\label{line:solution}
    \textbf{Return } Mechanism $\commech (\datavec) = \comb \datavec + N(\vec{0}, \comvar)$\;
    \caption{CommonMechanism($\mech_1, \mech_2$)}\label{alg:commech}
\end{algorithm}

\begin{algorithm}
   \DontPrintSemicolon
    \KwIn{Linear Gaussian Mechanism $\mech$ with query matrix $\bmat_{orig}$ and covariance matrix $\covar_{orig}$.}
    $\mat{X} \gets \bmat_{orig}^T \covar_{orig}^{-1} \bmat_{orig}$\tcp*{Privacy cost matrix}
    Use eigenvalue decomposition to represent $\mat{X} = \lambda_1 \vec{v}_1 \vec{v}_1^T + \cdots + \lambda_{\brank} \vec{v}_{\brank} \vec{v}_{\brank}^T + \cdots + \lambda_{\dimsize} \vec{v}_{\dimsize} \vec{v}_{\dimsize}^T$, where $\lambda_1, \cdots, \lambda_{\brank} > 0$, $\lambda_{{\brank}+1} = \cdots = \lambda_{\dimsize} = 0$\;
    $\bmat \gets  [\sqrt{\lambda_1} \vec{v}_{1}, \cdots, \sqrt{\lambda_{\brank}} \vec{v}_{\brank}]^T $ \tcp{$\bmat$ is matrix sqrt of $\mat{X}$}\label{line:stand:b}
    \textbf{Return } $\bmat$\;
    \caption{Standardization($\mech$)}\label{alg:standardize}
\end{algorithm}

First, it is easier to work with mechanisms $\mech_1$ and $\mech_2$ when their corresponding matrices $\bmat_1$ and $\bmat_2$ have linearly independent rows, and when the covariance matrices are the identity matrix. Thus we first perform a standardization step (Lines \ref{line:standard1}, \ref{line:standard2}) by calling Algorithm \ref{alg:standardize}, which rewrites $\mech_1$ (resp., $\mech_2)$ into an equivalent mechanism (Definition \ref{def:equal}) whose query matrix has linearly independent rows and the covariance matrix is the identity.

\begin{theoremEnd}[category=section5]{lemma}\label{lem:standardize}
Let $\mech_{orig}$ be a linear Gaussian mechanism with query matrix $\bmat_{orig}$ and covariance matrix $\covar_{orig}$. Let $\mech$ be a linear Gaussian mechanism with identity covariance and  query matrix $\bmat$ obtained by running Algorithm \ref{alg:standardize} on $\mech_{orig}$. Then $\mech_{orig}$ and $\mech$ are equivalent.
\end{theoremEnd}
\begin{proofEnd}
Using the eigenvalue decomposition of Algorithm \ref{alg:standardize}, we have the privacy cost matrix of $\mech_{orig}$ is  $\lambda_1 \vec{v}_1 \vec{v}_1^T + \cdots + \lambda_{\brank} \vec{v}_{\brank} \vec{v}_{\brank}^T + \cdots + \lambda_{\dimsize} \vec{v}_{\dimsize} \vec{v}_{\dimsize}^T$ which equals  $\lambda_1 \vec{v}_1 \vec{v}_1^T + \cdots + \lambda_{\brank} \vec{v}_{\brank} \vec{v}_{\brank}^T$ since the eigenvalues $\lambda_{k+1},\dots, \lambda_{\dimsize}$ are all 0. But that is the same as  $\bmat^T \mat{I}^{-1}\bmat$, the privacy cost matrix of $\mech$, due Line \ref{line:stand:b} in the algorithm,   and so they are equivalent by Theorem \ref{thm:answer}.
\end{proofEnd}

\subsubsection{An Optimization Problem for the Common Mechanism.}
Computing a maximal common mechanism $\commech$ requires finding a query matrix $\comb$ and covariance matrix $\comvar$. The following theorem allows us to select $\comb$ easily.

\begin{theoremEnd}[category=section5, all end]{lemma}\label{lem:basis}
Let $\mech(\datavec)=\bmat\datavec + N(\vec{0}, \covar)$ be a linear Gaussian mechanism. Let $\bmat^\prime$ be a matrix with linearly independent rows such that $\rowspace(\bmat) = \rowspace(\bmat^\prime)$. Then there exists a mechanism that is equivalent to $\mech$ and uses the query matrix $\bmat^\prime$.
\end{theoremEnd}
\begin{proofEnd}
Let $r$ be the number of rows of $\bmat$ (so it is a $r\times\dimsize$) matrix. This means that $\covar$ has size $r\times r$. Let $r^\prime$ be the number of rows of $\bmat^prime$.

Define $\mat{S}$ to be a $r\times r$ symmetric matrix square root of $\covar$ (which is possible because $\covar$ is positive semi-definite). Let $\mat{A}$ be a $r\times r^\prime$ matrix such that $\bmat = \mat{A}\bmat^\prime$ (which is possible because all the rows of $\bmat$ are in the row space of $\bmat^\prime)$. Note that the rank of $\bmat$ and $\bmat^\prime$ is the same, which equals $r^\prime$ and is the number of rows of $\bmat^\prime$. This means that the rank of $\mat{A}$ is also $r^\prime$.

Now consider the matrix $\mat{A}^{T}\covar^{-1}\mat{A}$ which is a symmetric positive semidefinite matrix of dimension $r^\prime \times r^\prime$. Since $\mat{A}$ has rank $r^\prime$ and $\covar$ has rank $r\geq r^\prime$, this matrix is invertible and therefore positive definite.  So we can define $\covar^\prime=\left(\mat{A}^{T}\covar^{-1}\mat{A}\right)^{-1}$.

Consider the mechanism $\mech^\prime(\datavec)=\bmat^\prime\datavec + N(\vec{0},\covar^\prime)$. We have
\begin{align*}
(\bmat^\prime)^T (\covar^\prime)^{-1}\bmat^\prime &= (\bmat^\prime)^T \mat{A}^{T}\covar^{-1}\mat{A}\bmat^\prime\\
&=\bmat^T \covar^{-1}\bmat^\prime
\end{align*}
and therefore $\mech$ and $\mech^\prime$ are equivalent by Theorem \ref{thm:answer} and Definition \ref{def:equal}.
\end{proofEnd}

\begin{theoremEnd}[category=section5]{theorem}\label{thm:chooseb}
Let $\mech_1(\datavec)=\bmat_1\datavec + N(\vec{0},\covar_1)$ and $\mech_2(\datavec)=\bmat_2\datavec + N(\vec{0}, \covar_2)$ be linear Gaussian mechanisms.  
\begin{itemize}[leftmargin=*]
\item If $\mech_c(\datavec)=\bmat_c\datavec + N(\vec{0},\covar_c)$ is common to $\mech_1$ and $\mech_2$ (i.e., $\mech_c$ is answerable from $\mech_1$ and also answerable from $\mech_2$) then $\rowspace(\bmat_c)\subseteq\rowspace(\bmat_1)\cap\rowspace(\bmat_2)$.
\item If $\mech_c$ is \emph{maximally} common then $\rowspace(\bmat_c)=$\\$\rowspace(\bmat_1)\cap\rowspace(\bmat_2)$
\item The choice for basis of $\rowspace(\bmat_1)\cap\rowspace(\bmat_2)$ does not matter. If $\mech_c$ is maximally common and if $\comb\neq \bmat_c$ is any matrix whose rows form a linearly independent basis, then there exists a common mechanism that is equivalent to $\mech_c$ and has query matrix $\comb$.
\end{itemize}
\end{theoremEnd}
\begin{proofEnd}
\textbf{First Item}.
In order for $\mech_c$ to be  common to $\mech_1$ and $\mech_2$, by Lemma \ref{thm:answer}, there must exist matrices $\mat{A}_1$, $\mat{A}_2$, such that
\begin{align*}
    \bmat_c &= \mat{A}_1 \bmat_1 \\
    \bmat_c &= \mat{A}_2 \bmat_2 
\end{align*}
and therefore $\rowspace(\bmat_c)\subseteq\rowspace(\bmat_1)\cap\rowspace(\bmat_2)$.

\textbf{Second Item}. 
If $\mech_c$ is maximally common but $\rowspace(\bmat_c)\subset\rowspace(\bmat_1)\cap\rowspace(\bmat_2)$ then there exists a unit (column) vector $\vec{v}\in \rowspace(\bmat_2)$ that is orthogonal to $\rowspace(\bmat_c)$. Define $\frac{1}{\sigma^2} =\min(\vec{v}^T\bmat^T_1\covar_1^{-1}\bmat_1)\vec{v},~\vec{v}^T\bmat^T_2\covar_2^{-1}\bmat_2\vec{v})$ and note that $\frac{1}{\sigma^2}$ is nonzero because $\vec{v}$ is in the row space of these matrices and the covariance matrices are nonsingular.
Define
\begin{align*}
\bmat^\prime &= 
\left[\begin{smallmatrix}\bmat_c\\
\vec{v}^T\end{smallmatrix}\right] \\
\covar^\prime &= \left[\begin{smallmatrix}\covar_c & \mat{0}\\\mat{0} & \sigma^2\end{smallmatrix}\right]
\end{align*}
and consider the mechanism $\mech^\prime(\datavec) =\bmat^\prime\datavec + N(\vec{0},\covar^\prime)$. Clearly, $\mech_c$ is answerable from $\mech^\prime$ simply by dropping the last query answer from $\mech$ (corresponding to the last row of $\bmat^\prime$). On the other hand, it is easy to see that $\mech^\prime$ is not answerable from $\mech_c$.

By Lemma \ref{lem:basis} we can, without loss of generality, assume that  $\bmat_1$ has linearly independent rows, that one of the rows is $\vec{v}^T$ and the rest of the rows of $\bmat_1$ are orthogonal to $\vec{v}^T$ (since $\vec{v}^T$ is in the row space of $\bmat_1$). The same can be assumed of $\bmat_2$, so that we can write $\bmat_1 = \left[\begin{smallmatrix}\bmat_1^\circ\\\vec{v}^T\end{smallmatrix}\right]$ for some $\bmat_1^\circ$ whose rows are orthogonal to $\vec{v}^T$ and similarly $\bmat_2=\left[\begin{smallmatrix}\bmat_2^\circ\\\vec{v}^T\end{smallmatrix}\right]$

Now let $\vec{w}$ be any vector orthogonal to $\vec{v}$. Since $\mech_c$ is common to $\mech_1$ and $\mech_2$, then by Theorem \ref{thm:answer}, we must have:
\begin{align}
0 &\leq \vec{w}^T(\bmat_1^T\covar_1^{-1}\bmat_1 - \bmat_c^T\covar_c^{-1}\bmat_c)\vec{w}^T\nonumber\\
&\vec{w}^T\left(\left[\begin{smallmatrix}\bmat_1^\circ\\\vec{v}^T\end{smallmatrix}\right]^T\covar_1^{-1}\left[\begin{smallmatrix}\bmat_1^\circ\\\vec{v}^T\end{smallmatrix}\right] - \bmat_c^T\covar_c^{-1}\bmat_c\right)\vec{w}^T\nonumber\\
&= \vec{w}^T\left(\left[\begin{smallmatrix}\bmat_1^\circ\\\vec{0}^T\end{smallmatrix}\right]^T\covar_1^{-1}\left[\begin{smallmatrix}\bmat_1^\circ\\\vec{0}^T\end{smallmatrix}\right] - \bmat_c^T\covar_c^{-1}\bmat_c\right)\vec{w}^T\label{eq:rowspace1}\\
0 &\leq \vec{w}^T(\bmat_2^T\covar_2^{-1}\bmat_2 - \bmat_c^T\covar_c^{-1}\bmat_c)\vec{w}^T\nonumber\\
&= \vec{w}^T\left(\left[\begin{smallmatrix}\bmat_2^\circ\\\vec{0}^T\end{smallmatrix}\right]^T\covar_2^{-1}\left[\begin{smallmatrix}\bmat_2^\circ\\\vec{0}^T\end{smallmatrix}\right] - \bmat_c^T\covar_c^{-1}\bmat_c\right)\vec{w}^T\label{eq:rowspace2}
\end{align}
Also, by construction of $\sigma^2$, we have:
\begin{align}
\vec{v}^T\left(\bmat_1^T\covar_1^{-1}\bmat_1 - \frac{1}{\sigma^2}\vec{v}^T\vec{v}\right)\vec{v}^T &= \vec{v}^T\left(\bmat_1^T\covar_1^{-1}\bmat_1\right)\vec{v} - \frac{1}{\sigma^2}\geq 0
\label{eq:rowspace3}\\
\vec{v}^T\left(\bmat_2^T\covar_2^{-1}\bmat_2 - \frac{1}{\sigma^2}\vec{v}^T\vec{v}\right)\vec{v}^T &\geq 0
\label{eq:rowspace4}
\end{align}
Now we show that $\mech^\prime$ is also common to $\mech_1$ and $\mech_2$.

Any vector $\vec{z}$ can be represented as a linear combination $\vec{w} + b \vec{v}$ where $\vec{w}$ is a vector that is orthogonal to $\vec{v}$ (and the choice of $b,\vec{w}$ depends on $\vec{z}$).
\begin{align*}
\lefteqn{
\vec{z}^T \left(\bmat_1^T \covar_1^{-1}\bmat_1 - (\bmat^\prime)^T(\covar^\prime)^{-1}(\bmat^\prime)\right)\vec{z}}\\
&=
(\vec{w} + b \vec{v})^T \left(\bmat_1^T \covar_1^{-1}\bmat_1 - (\bmat^\prime)^T(\covar^\prime)^{-1}(\bmat^\prime)\right)(\vec{w} + b \vec{v})\\
&=
(\vec{w} + b \vec{v})^T \left(\bmat_1^T \covar_1^{-1}\bmat_1 - \left[\begin{smallmatrix}\bmat_c\\
\vec{v}^T\end{smallmatrix}\right]^T
\left[\begin{smallmatrix}\covar_c & \mat{0}\\\mat{0} & \sigma^2\end{smallmatrix}\right]^{-1}
\left[\begin{smallmatrix}\bmat_c\\
\vec{v}^T\end{smallmatrix}\right]
\right)(\vec{w} + b \vec{v})\\
&=
(\vec{w} + b \vec{v})^T \left(\bmat_1^T \covar_1^{-1}\bmat_1 - \left[\begin{smallmatrix}\bmat_c^T, &~~
\vec{v}\end{smallmatrix}\right]
\left[\begin{smallmatrix}\covar_c^{-1} & \mat{0}\\\mat{0} & 1/\sigma^2\end{smallmatrix}\right]
\left[\begin{smallmatrix}\bmat_c\\
\vec{v}^T\end{smallmatrix}\right]
\right)(\vec{w} + b \vec{v})\\
&=(\vec{w} + b \vec{v})^T \left(\bmat_1^T \covar_1^{-1}\bmat_1 - 
\bmat_c^T \covar_c^{-1}\bmat_c - \frac{1}{\sigma^2}\vec{v}\vec{v}^T
\right)(\vec{w} + b \vec{v})\\
&= \vec{w}^T\left(\bmat_1^T\covar_1^{-1}\bmat_1 - \bmat_c^T\covar_c^{-1}\bmat_c\right)\vec{w}^T + b^2\vec{v}^T\left(\bmat_1^T\covar_1^{-1}\bmat_1 - \frac{1}{\sigma^2}\vec{v}^T\vec{v}\right)\vec{v}^T\\
&\quad+ 2b\vec{w}^T\left(\bmat_1^T\covar_1^{-1}\bmat_1\right)\vec{v}\\
\intertext{Since $\vec{v}^T$ is orthogonal to $\vec{w}$ and the rows of $\bmat_c$}
&= \vec{w}^T\left(\bmat_1^T\covar_1^{-1}\bmat_1 - \bmat_c^T\covar_c^{-1}\bmat_c\right)\vec{w}^T + b^2\vec{v}^T\left(\bmat_1^T\covar_1^{-1}\bmat_1 - \frac{1}{\sigma^2}\vec{v}^T\vec{v}\right)\vec{v}^T\\
&\quad+ 2b\vec{w}^T\left(\left[\begin{smallmatrix}(\bmat_1^\circ)^T,~~ &\vec{v}\end{smallmatrix}\right]\covar_1^{-1}\left[\begin{smallmatrix}\bmat_1^\circ\\\vec{v}^T\end{smallmatrix}\right]\right)\vec{v}\\
&= \vec{w}^T\left(\bmat_1^T\covar_1^{-1}\bmat_1 - \bmat_c^T\covar_c^{-1}\bmat_c\right)\vec{w}^T + b^2\vec{v}^T\left(\bmat_1^T\covar_1^{-1}\bmat_1 - \frac{1}{\sigma^2}\vec{v}^T\vec{v}\right)\vec{v}^T\\
&\quad+ 2b\vec{w}^T\left(\left[\begin{smallmatrix}(\bmat_1^\circ)^T,~~ &0\end{smallmatrix}\right]\covar_1^{-1}\left[\begin{smallmatrix}\mat{0}\\\vec{v}^T\end{smallmatrix}\right]\right)\vec{v}
\intertext{Since $\vec{v}^T$ is orthogonal to $\vec{w}$ and the rows of $\mat{B}_1^\circ$}
&= \vec{w}^T\left(\bmat_1^T\covar_1^{-1}\bmat_1 - \bmat_c^T\covar_c^{-1}\bmat_c\right)\vec{w}^T + b^2\vec{v}^T\left(\bmat_1^T\covar_1^{-1}\bmat_1 - \frac{1}{\sigma^2}\vec{v}^T\vec{v}\right)\vec{v}^T\\
\end{align*}
and both terms are nonnegative because of Equations \ref{eq:rowspace1} and \ref{eq:rowspace3}. A similar result holds for $\bmat_2$. Thus, by Theorem \ref{thm:answer}, $\mech^\prime$ is common to $\mech_1$ and $\mech_2$.

Since $\mech_c$ is answerable from $\mech^\prime$ (but not vice versa) then $\mech_c$ is not maximal. This contradiction arose from the assumption that the row space of $\bmat_c$ was a strict subset of the intersection of the row spaces of $\bmat_1$ and $\bmat_2$.

\textbf{Third item}. This follows directly from Lemma \ref{lem:basis}.

\end{proofEnd}
Thus we set $\comb$ to be a matrix whose rows form a basis for $\rowspace(\bmat_1)\cap\rowspace(\bmat_2)$ in Line \ref{line:basis} in Algorithm \ref{alg:commech}. This can be done in multiple ways, such as using the Zassenhaus algorithm \cite{zassenhaus} or eigendecompositions \cite{batta}. Then we use the following theorem to set up an optimization problem for finding a covariance matrix.
\begin{theoremEnd}[category=section5]{theorem}
Let $\mech_1(\datavec)=\bmat_1\datavec + N(\vec{0},\mat{I})$ and $\mech_2(\datavec)=\bmat_2\datavec + N(\vec{0}, \mat{I})$ be  linear Gaussian mechanisms that are \underline{standardized} (i.e., produced by Algorithm \ref{alg:standardize}).  Let $\comb$ be a matrix whose rows are linearly independent and spans $\rowspace(\bmat_1)\cap\rowspace(\bmat_2)$. Then one can obtain a maximally common mechanism by using the $\comvar$ that optimizes the following problem (here $\dagger$ represents the Moore-Penrose Pseudoinverse operation):
\begin{align}
\label{prob:min-trace}
    \comvar \gets\min_{\covar} ~ trace(\covar) ~
    s.t. ~& \covar ~ \rloewner ~\comb \bmat^\dagger_1  (\bmat^\dagger_1)^T\comb^T \\
    \nonumber
    & \covar ~ \rloewner ~\comb \bmat^\dagger_2  (\bmat^\dagger_2)^T\comb^T 
    \nonumber
\end{align}
\end{theoremEnd}
\begin{proofEnd}
In order for a mechanism $\commech(\datavec) =\comb\datavec + N(\vec{0}, \comvar)$ to be answerable from $\mech_1$ and $\mech_2$, by Lemma \ref{def:answer}, there must exist  matrices $\mat{A}_1$ and $\mat{A}_2$ such that 
\begin{align*}
    \comb &= \mat{A}_1 \bmat_1 \\
    \comb &= \mat{A}_2 \bmat_2 \\
    \comvar & \rloewner \mat{A}_1  \mat{A}_1^T \\
    \comvar & \rloewner \mat{A}_2  \mat{A}_2^T
\end{align*}
since $\covar_1=\mat{I}$ and $\covar_2=\mat{I}$. Since the rows of $\bmat_1$ are linearly independent (and similarly for $\bmat_2$), the matrices $\mat{A}_1$ and $\mat{A}_2$ are unique and can be obtained by multiplying by the right inverses (and therefore Moore-Penrose Pseudoinverses) $\bmat_1^\dagger, \bmat_2^\dagger$, so that we can set:
\begin{align*}
\mat{A}_1 &= \comb \bmat_1^\dagger\\
\mat{A}_2 &= \comb \bmat_2^\dagger
\end{align*}
Then we can minimize $\comvar$ over the Loewner partial order subject to the constraints $\comvar \rloewner \mat{A}_1  \mat{A}_1^T$  and     $\comvar  \rloewner \mat{A}_2  \mat{A}_2^T$. This will ensure that we get a maximally common mechanism because (1) any other common mechanism can be rewritten so that its query matrix is also $\comb$ (by Lemma \ref{lem:basis}) and its covariance matrix $\comvar^\prime$ must also be larger than $\mat{A}_1  \mat{A}_1^T$ and $\mat{A}_1  \mat{A}_1^T$ in the Loewner order; (2) if this other mechanism with covariance $\comvar^\prime$ answers the one with covariance $\comvar$ then by Lemma \ref{lem:answer} we must have $\comvar\rloewner\comvar^\prime$, (3) but by minimality of $\comvar$, we therefore have $\comvar=\comvar^\prime$. 

However, the Loewner partial order means that there could be multiple minimal solutions, and so we are free to pick one by using an objective function that respects the Loewner partial order. One such objective function is $\trace(\covar)$ and one can easily see that $\covar_a \loewner \covar_b$ implies $\trace(\covar_a)\leq \trace(\covar_b)$ since:
\begin{align*}
\trace(\covar_b-\covar_a) &= \sum_i \vec{e}_i^T (\covar_b-\covar_a) \vec{e}_i
\intertext{where $\vec{e}_i$ has a 1 in position $i$ and 0 everywhere else}
&\geq 0 \text{ by the Loewner order on $\covar_a$ and $\covar_b$}
\end{align*}
Putting all these facts together results in the optimization problem in Equation \ref{prob:min-trace}.
\end{proofEnd}

To find the maximally common mechanism of 3 or more mechanisms, simply add an additional constraint for each mechanism in the optimization for Equation \ref{prob:min-trace}.
However, when dealing with just two mechanisms, the optimization problem in Equation \ref{prob:min-trace} has a symbolic solution that is  used in Line \ref{line:solution} in Algorithm \ref{alg:commech}. This kind of matrix optimization  was studied and solved by Stott \cite{stott2017minimal,stott2016maximal,stott2017universite}:

\begin{theoremEnd}[category=section5]{theorem}\label{thm:solution}[Stott \cite{stott2017minimal,stott2016maximal,stott2017universite}]
The solution to the optimization problem in Equation \ref{prob:min-trace} is $\comvar = \frac{\mat{A}_{1}  \mat{A}_{1}^T + \mat{A}_{2}  \mat{A}_{2}^T}{2} + \frac{|\mat{A}_{2}  \mat{A}_{2}^T - \mat{A}_{1}  \mat{A}_{1}^T|}{2}$, where $\mat{A}_1=\comb\bmat_1^\dagger$, $\mat{A}_2=\comb\bmat_2^\dagger$, and $|\cdot|$ is the operator that replaces negative eigenvalues with positive eigenvalues (i.e., if the eigendecomposition of $\mat{V}=\mat{P}^T Diag(\vec{\lambda})\mat{P}$ then $|\mat{V}|=\mat{P}^T Diag(|\vec{\lambda}|)\mat{P}$).
\end{theoremEnd}

\subsection{Computing Residual Mechanisms}\label{subsec:makeresidual}

\begin{algorithm}
   \DontPrintSemicolon
    \KwIn{Intended Mechanism $\mech_i(\datavec)=\bmat_i \datavec + N(\vec{0}, \covar_i)$. \\ Common mechanism $\commech(\datavec)=\comb\datavec + N(\vec{0},\comvar)$}
    $\mat{X} \gets \bmat_i^T \covar_i^{-1} \bmat_i - \comb^T \comvar^{-1} \comb$\; \label{line:pcostmat}
    Use eigenvalue decomposition to represent $\mat{X} = \lambda_1 \vec{v}_1 \vec{v}_1^T + \cdots + \lambda_{\brank} \vec{v}_{\brank} \vec{v}_{\brank}^T + \cdots + \lambda_{\dimsize} \vec{v}_{\dimsize} \vec{v}_{\dimsize}^T$, where $\lambda_1, \cdots, \lambda_{\brank} > 0$, $\lambda_{{\brank}+1} = \cdots = \lambda_{\dimsize} = 0$\;
    $\bmat^\prime_i \gets  [\sqrt{\lambda_1} \vec{v}_{1}, \cdots, \sqrt{\lambda_{\brank}} \vec{v}_{\brank}]^T $\tcp*{$\bmat^\prime_i$ is matrix sqrt of $\mat{X}$}
    $\covar^\prime_i \gets \mat{I}_*$\;
    \textbf{Return } Residual mechanism $\mech' (\datavec) = \bmat^\prime_i \datavec + N(\vec{0}, \covar^\prime_i)$ \;
    \caption{ResidualMechanism($\mech_i, \mech_*$)}\label{alg:residual}
\end{algorithm}

Given a mechanism $\mech_1(\datavec)=\bmat_1\datavec + N(\vec{0},\covar)$ and a maximally common mechanism $\commech(\datavec)=\comb\datavec + N(\vec{0},\comvar)$, computing the residual mechanism $\mech^\prime_1$ is greatly simplified by Lemma \ref{lem:residual}. One simply needs to find a $\bmat_1^\prime$ and $\covar_1^\prime$ that satisfies:
$$(\bmat_1^\prime)^T (\covar_1^\prime)^{-1} \bmat_1^\prime ~= ~\bmat_1^T \covar_1^{-1} \bmat_1 ~-~\comb^T \comvar^{-1}\comb. 
$$
This operation is performed by Algorithm \ref{alg:residual}.

\subsection{Recreating the target mechanisms}\label{subsec:recreate}
Once one has obtained the output $\outp_*$ of the common mechanism $\commech$, one would run the residual mechanism $\mech^\prime_i$ of the chosen mechanism $\mech_i$ (i.e., $\mech_1$ or $\mech_2$) to obtain the output $\outp_i^\prime$. 

The next step is to use $\outp_*$ and $\outp_i^\prime$ to provide the same answer $\mech_i$ would have produced. It is a postprocessing step and does not consume any privacy budget. It is a linear function of the vectors $\outp_*$ and $\outp_i^\prime$, shown in Algorithm \ref{alg:recovery} and justified by Theorem \ref{thm:recovery}.

\begin{algorithm}
   \DontPrintSemicolon
   \KwIn{$\outp_*$: output of the common  mechanism \\
   \phantom{\textbf{Input: }}$\outp^\prime_i$: output of a residual mechanism.\\
      \phantom{\textbf{Input: }} $\comb, \comvar$: query and cov. matrices of common mech.\\
   \phantom{\textbf{Input: }}$\bmat_i, \covar_i$: matrices for target mech. (e.g., $\mech_1$ or $\mech_2$).\\
   \phantom{\textbf{Input: }}$\bmat_i^\prime. \covar_i^\prime$: query and cov. matrices of residual mech.}
   $\covar_i^{1/2}\gets$ symmetric matrix sqrt of $\covar_i$\;
   $\mat{W} \gets \bmat_i^T (\covar_i^{1/2})^{-1}$\;
   $\mat{A}_* \gets \covar_i^{1/2} \mat{W}^\dagger \comb^T\comvar^{-1}$\;
   $\mat{A}_i^\prime \gets \covar_i^{1/2} \mat{W}^\dagger \bmat_i^{\prime T}(\covar_i^\prime)^{-1}$\;
   \Return{$\mat{A}_* \outp_* + \mat{A}_i^\prime \outp^\prime_i$}
    \caption{Recreate($\outp_*, \outp'$)}\label{alg:recovery}
\end{algorithm}

\begin{theoremEnd}[category=section5]{theorem}\label{thm:recovery}
Let $\mech_1(\datavec)=\bmat_1\datavec + N(\vec{0},\covar_1)$ and $\mech_2(\datavec)=\bmat_2\datavec + N(\vec{0},\covar_2)$ be linear Gaussian mechanisms. Let $\commech(\datavec) = \comb\datavec + N(\vec{0},\comvar)$ be their maximally common mechanism and let $\mech^\prime_1(\datavec) = \bmat^\prime_1\datavec + N(\vec{0},\covar_1^\prime)$ be the residual mechanism for $\mech_1$. Define:
\begin{itemize}
\item $\covar_1^{1/2}$ to be the symmetric matrix square root of $\covar_1$, 
\item $\covar_1^{-1/2}$ to be the inverse of  $\covar_1^{1/2}$,
\item $\mat{A}_* = \covar_1^{1/2}(\bmat^T_1 \covar_1^{-1/2})^\dagger\comb^T\comvar^{-1}$, where $\dagger$ is the Moore-Penrose pseudo-inverse,
\item $\mat{A}_1^\prime = \covar_1^{1/2}(\bmat^T_1 \covar_1^{-1/2})^\dagger\bmat_1^{\prime T}(\covar_1^\prime)^{-1}$.
\end{itemize}
Then $    \mech_1(\datavec) = \mat{A}_* \commech(\datavec) + \mat{A}_1^{\prime} \mech_1^{\prime}(\datavec) + N(\vec{0}, \covar_1 - \mat{A}_*\comvar\mat{A}_*^T -  \mat{A}_1^{\prime}\covar_1^\prime\mat{A}_1^{\prime T})$ and $\mech_1(\datavec)$ is equivalent to $\mat{A}_* \commech(\datavec) + \mat{A}_1^{\prime} \mech_1^{\prime}(\datavec)$.
\end{theoremEnd}

\begin{proofEnd}
\textbf{Step 1}.
First we need to show that for any $\datavec$, $E[\mech_1(\datavec)] = \mat{A}_*E[\commech(\datavec)] + \mat{A}_1^\prime E[\mech_1^\prime(\datavec)]$, which is the same as showing $\bmat_1 = \mat{A}_*\comb + \mat{A}_1^\prime \bmat_1^\prime$.

Express $\bmat^T_1 \covar_1^{-1/2}$ in terms of its singular value decomposition:
$$\bmat^T_1 \covar_1^{-1/2} = \mat{U} \mat{D} \mat{V}^T$$
where $\mat{D}$ is an invertible  $\numquery \times \numquery$ diagonal matrix (for some $\numquery$), $\mat{U}$ is an $\dimsize \times\numquery$ matrix with orthogonal columns such that $\mat{U}^T \mat{U} = \mat{I}_{\numquery}$, and $\mat{V}$ is a $\dimsize\times\numquery$ orthogonal matrix, so that $\mat{V}^T \mat{V} = \mat{I}_{\numquery}$. Then we have:
\begin{align*}
(\bmat^T_1 \covar_1^{-1/2})^\dagger = \mat{V} \mat{D}^{-1} \mat{U}^T
\end{align*}
 Recall that both common and residual mechanisms are constructed so that:
$\comb^T \comvar^{-1}\comb ~+~ (\bmat_1^\prime)^T (\covar_1^\prime)^{-1} \bmat_1^\prime ~= ~\bmat_1^T \covar_1^{-1} \bmat_1$, we have
\begin{align*}
    \mat{A}_* \comb + \mat{A}_1^{\prime} \bmat_1^{\prime} & = \covar_1^{1/2}(\bmat^T_1 \covar_1^{-1/2})^\dagger\left(\comb^T\comvar^{-1}\comb + \bmat_1^{\prime T}(\covar_1^\prime)^{-1}\bmat^\prime_1\right)\\
   &=\covar_1^{1/2}(\bmat^T_1 \covar_1^{-1/2})^\dagger\left( \bmat_1^{T}\covar_1^{-1}\bmat_1\right)\\
    &=\covar_1^{1/2}(\bmat^T_1 \covar_1^{-1/2})^\dagger\left( \bmat_1^{T}\covar_1^{-1/2}\right)\left(\covar_1^{-1/2}\bmat_1\right)\\
       &=\covar_1^{1/2} \mat{V}\mat{D}^{-1}\mat{U}^T \mat{U}\mat{D}\mat{V}^T\mat{V}\mat{D}\mat{U}^T\\
          &=\covar_1^{1/2} \mat{V}\mat{D}^{-1}\mat{D}\mat{D}\mat{U}^T\\
                   &=\covar_1^{1/2} \mat{V}\mat{D}\mat{U}^T\\
                   &=\covar_1^{1/2}\left(\covar_1^{-1/2}\bmat_1\right)\\
                   &=\bmat_1
\end{align*}

\noindent\textbf{Step 2:} Now, by Lemma \ref{lem:answer} we need to show that the covariance matrix of  $\mat{A}_*E[\commech(\datavec)] + \mat{A}_1^\prime E[\mech_1^\prime(\datavec)]$ is $\loewner ~\covar_1$ in the Loewner partial order. This covariance matrix is:
\begin{align*}
\lefteqn{\mat{A}_* \comvar \mat{A}_*^T + \mat{A}_1^\prime \covar_1^\prime \mat{A}_1^{\prime T}}\\
&= 
\covar_1^{1/2}(\bmat^T_1 \covar_1^{-1/2})^\dagger
\Big(
\comb^{ T}\comvar^{-1}\comvar\comvar^{-1}\comb\\
&\phantom{=\covar_1^{1/2}(\bmat^T_1 \covar_1^{-1/2})^\dagger}+
\bmat_1^{\prime T}(\covar_1^\prime)^{-1}\covar_1^\prime (\covar_1^\prime)^{-1}\bmat_1^{\prime}
\Big)
(\bmat^T_1 \covar_1^{-1/2})^{\dagger T}\covar_1^{1/2}\\
&=
\covar_1^{1/2}(\bmat^T_1 \covar_1^{-1/2})^\dagger
\left(
\comb^{ T}\comvar^{-1}\comb
+
\bmat_1^{\prime T} (\covar_1^\prime)^{-1}\bmat_1^{\prime}
\right)
(\bmat^T_1 \covar_1^{-1/2})^{\dagger T}\covar_1^{1/2}\\
&=
\covar_1^{1/2}(\bmat^T_1 \covar_1^{-1/2})^\dagger
\left(
\bmat_1^T \covar_1^{-1} \bmat_1
\right)
(\bmat^T_1 \covar_1^{-1/2})^{\dagger T}\covar_1^{1/2}\\
&=
\covar_1^{1/2}
\mat{V}\mat{D}^{-1}\mat{U}^T
\left(\mat{U}\mat{D}\mat{V}^T\mat{V}\mat{D}\mat{U}^T\right)
\mat{U}\mat{D}^{-1}\mat{V}^T
\covar_1^{1/2}\\
&=
\covar_1^{1/2}
\mat{V}\mat{D}^{-1}\mat{D}\mat{D}\mat{D}^{-1}\mat{V}^T
\covar_1^{1/2}\\
&=\covar_1^{1/2}
\mat{V}\mat{V}^T
\covar_1^{1/2}\\
&=\covar_1^{1/2}\left(
\sum_{i=1}^{\numquery}\vec{v}_i\mat{v}_i^T\right)
\covar_1^{1/2}\\
\intertext{(where the $\vec{v}_i$ are the orthogonal columns of $\mat{V}$)}
&\loewner \covar_1^{1/2}\left(
\sum_{i=1}^{\numquery}\vec{v}_i\mat{v}_i^T\right)
\covar_1^{1/2} + \covar_1^{1/2}\left(
\sum_{i=\numquery+1}^{\dimsize}\vec{v}^\prime_i\mat{v}_i^{\prime T}\right)
\covar_1^{1/2}\\
\intertext{(where the $\vec{v}_i^\prime$ are orthogonal and complete the bases for $\mathbb{R}^\dimsize$)}
&= \covar_1^{1/2}\covar_1^{1/2} = \covar_1
\end{align*}

\noindent\textbf{Step 3:} The equivalence claimed in the theorem follows from the fact that $\mech_1$ is answerable from $\mat{A}_* \commech + \mat{A}_1^{\prime} \mech_1^{\prime}$ (which was just proven) and $(\commech, \mech_1^\prime)$ is answerable from $\mech_1$, by definition of common and residual mechanism, and $\mat{A}_* \commech + \mat{A}_1^{\prime} \mech_1^{\prime}$ is answerable from $(\commech, \mech_1^\prime)$.
\end{proofEnd}

\section{Making decisions based on the common mechanism.}
\label{sec:decide}
We next consider how one could use the output $\outp_*$ of the common mechanism to decide whether to run the residual mechanism $\mech_1^\prime$ in order to do the analysis supported by $\mech_1$, or whether to run $\mech_2^\prime$ instead. In general, this would be user/application dependent, but we list some suggestions here. We first consider the general case where $\mech_1$ and $\mech_2$ are arbitrary linear Gaussian mechanisms in Section \ref{subsec:select:general} and nested analyses, such as  1-way vs. 2-way marginals, in Section \ref{subsec:select:nested}

\subsection{The General Case}\label{subsec:select:general}
One can use the answer $\outp_*$ of the common mechanism to estimate how much potential variability there can be in the data $\datavec$ and compare it to the amount of noise that $\mech_1$ or $\mech_2$ would add. One would then choose between them based on the potential signal-to-noise ratio. For example, if the amount of noise required by $\mech_1$ (resp., $\mech_2$) is larger than the range of potential values for $\bmat_1\datavec$ (resp., $\bmat_2\datavec$), then it would not be worthwhile to choose $\mech_1$ (resp., $\mech_2$).

Given the output $\outp_*$ of the common mechanism, one can do the following postprocessing to estimate a range of possible datasets that are compatible with $\outp_*$. 
First, one would generate a set of random vectors $\vec{c}_1,\dots, \vec{c}_r$. For each $\vec{c}_i$, one would solve the following optimization problem (in Equation \ref{prob:general}) to get an estimate $\widehat{\vec{x}}_i$ for a potential dataset, and then one would compute $\bmat_1\widehat{\vec{x}}_i$. By computing the variance of all of these values $\bmat_1\widehat{\vec{x}}_1, \dots, \bmat_1\widehat{\vec{x}}_r$ one would see the range of possible answers to the query matrix $\bmat_1$ of $\mech_1$, and by comparing it to the (total variance of the) covariance matrix $\covar_1$, one would determine whether the noise would mask these variations (meaning that nothing useful would be learned from a noisy answer).

\begin{align}
\label{prob:general}
    \min_{\widehat{\datavec}} ~& \vec{c}^T \bmat_1 \widehat{\datavec}\\
    \nonumber
    s.t. & ||\comvar^{-1/2}(\comb \widehat{\datavec} - \outp_{*})||_1 \leq 2 \\
    \nonumber
    & \widehat{\datavec} \succeq \vec{0}
\end{align}
The constraints in Equation \ref{prob:general} say that one is looking for a nonnegative vector of counts $\widehat{\vec{x}}$ that is consistent with the noisy answer to the common mechanism, up to 2 standard deviations. Any feasible solution is a candidate dataset. The random weight vector $\vec{c}$ causes a random feasible solution to be returned.

\subsection{Nested Analyses}\label{subsec:select:nested}
Another typical use-case occurs when analyses are nested. For example, an analyst may be interested in a primary analysis involving one-way marginals. However, there might also be interest in performing a secondary analysis using two-way marginals. This is an example of \emph{nesting} because one-way marginals can be computed from two-way marginals.

When an analyst is only allowed to obtain marginals that are noisy, an important decision must be made.
\begin{itemize}
\item If the analyst requests noisy one-way marginals via a mechanism $\mech_1(\datavec)=\bmat_1\datavec + N(\vec{0},\covar_1)$ then the analyst gets the most accurate noisy information needed for their primary analysis, but no secondary analysis can be performed.
\item If the analyst requests noisy two-way marginals via a mechanism $\mech_2(\datavec)=\bmat_2\datavec + N(\vec{0},\covar_2)$. This would allow the secondary analysis to proceed. Since noisy two-way marginals can be used to compute noisier one-way marginals (these one-way marginals will be noisier than in the first option), then the primary analysis can also be performed if there was enough signal-to-noise ratio (low relative error).
\end{itemize}
Here we define the signal-to-noise ratio (SNR) as the true (unknown) count of a cell divided by the standard deviation of the privacy noise applied to it.
Thus, if noisy two-way marginals have enough signal-to-noise ratio to perform the primary analysis, then
$\mech_2$ is preferable since $\mech_2$ also has additional uses. Otherwise, $\mech_1$ would be preferable. The goal is for the analyst to use the output of the common mechanism to estimate what the SNR of each marginal cell would be if $\mech_1$ or $\mech_2$ were used.

We say that the analysis supported by query matrix $\bmat_1$ (e.g., one-way marginals) is nested in the analysis supported by $\bmat_2$ (e.g., two-way marginals) if there exists a matrix $\mat{A}$ such that $\bmat_1  = \mat{A}\bmat_2$. In this case, $\mat{A}\mech_2(\datavec)$ (i.e, multiplying the output of $\mech_2$ by $\mat{A}$) is a noisier version of $\mech_1$. We say that $\bmat_1$ is the primary analysis and $\bmat_2$ is the secondary analysis.

Now, the common mechanism $\commech$ of $\mech_1$ and $\mech_2$ can provide a noisy answer to any query in the intersection of the row spaces of $\bmat_1$ and $\bmat_2$ (by Theorem \ref{thm:chooseb}) and so there exists a matrix $\mat{A}_*$ such that $E[\mech_1(\data)]=E[\mat{A}_*\commech(\data)]$, which means that $\mat{A}_*\commech$ is also a noisier version of $\mech_1$, and the variance of the $i^\text{th}$ query is the $i^\text{th}$ diagonal element of $\mat{A}_*\comvar\mat{A}_*^T$, denoted by $(\mat{A}_*\comvar\mat{A}_*^T)[i,i]$.  Thus the output $\outp_*$ of $\commech$ can be used to estimate the signal to noise ratio of using mechanism $\mech_2$ to do the nested primary analysis as follows (all of this is a postprocessing of $\outp_*$):

\begin{itemize}
\item The quantity $L_i = (\mat{A}_*\outp_*)[i] - 3\sqrt{(\mat{A}_*\comvar\mat{A}_*^T)[i,i]}$ is a 3 sigma lower confidence interval for the true value of the $i^\text{th}$ query in  $\bmat_1\datavec$. The quantity $U_i = (\mat{A}_*\outp_*)[i] + 3\sqrt{(\mat{A}_*\comvar\mat{A}_*^T)[i,i]}$ is the corresponding upper confidence interval.
\item The quantity $SNR\_Lower^{(1)}_{i} = \frac{L_i}{\sqrt{\covar_1[i,i]}}$ is a lower bound on the expected signal-to-noise ratio of using $\mech_1$ to get a noisy answer to the  $i^\text{th}$ query in  $\bmat_1\datavec$. Similarly, $SNR\_Upper^{(1)}_{i}=\frac{U_i}{\sqrt{\covar_1[i,i]}}$ would be an upper bound on the signal-to-noise ratio.
\item The quantity $SNR\_Lower^{(2)}_{i} = \frac{L_i}{\sqrt{(\mat{A}\covar_2\mat{A}^T)[i,i]}}$ (resp.,\\ $SNR\_Upper^{(2)}_{i}=\frac{U_i}{\sqrt{(\mat{A}\covar_2\mat{A}^T)[i,i]}}$) is a lower (resp., upper) bound on the expected signal-to-noise ratio of using $\mech_2$ to get a noisy answer to the  $i^\text{th}$ query in  $\bmat_1\datavec$ (the nested analysis).
\end{itemize}

These quantities can be used in a variety of ways. For example, the user may want at least $x\%$ of the queries of  $\bmat_1$ to have SNR above $y$. In this case, if $x\%$ of the $SNR\_Lower^{(2)}_i$ values are $\geq y$,  then $\mech_2$ is good enough for the primary analysis and also provides an opportunity to perform the secondary analysis. Thus, the user should decide to run the residual mechanism $\mech^\prime_2$ to get the answer to $\mech_2$.

Other possibilities also exist. If too many of the $SNR\_Upper^{(1)}_i$ values are below the desired signal-to-noise ratio $y$, then even $\mech_1$ is not accurate enough for the primary analysis and the analyst can stop here, without using any further privacy budget beyond what $\commech$ had cost. On the other hand, if the SNR bound $y$ is between $SNR\_Lower^{(2)}_i$ and $SNR\_Upper^{(2)}_i$ for many $i$, then the analyst also has the option of using more privacy budget to help make the decision between $\mech_1$ and $\mech_2$, or may opt for $\mech_1$ just to be safe.

\section{Experiments}\label{sec:experiments}

We conduct experiments to examine how well the common mechanism can guide the analyst when choosing between nested analyses, as described in Section \ref{subsec:select:nested}. We use three real datasets and several applications, one of which (Section \ref{sec:exp:census}) is a comparison to an algorithm that the Census Bureau will use as part of the Detailed Demographic and Housing Characteristics (DHC) data release \cite{dhc}.

\subsection{Datasets}\label{sec:data}
We use three datasets: \textbf{HispRace}, \textbf{AgeGender}, and \textbf{Brazil} \cite{ipums}. The \textbf{HispRace} dataset is extracted from the 2010 Summary File 1 (SF1) tabulations P4 and P5 \cite{census2010}. For each of the $6,257,947$ occupied census blocks  in the dataset, a record has 7 binary race and ethnicity attributes for a domain size of $2^7=128$ for each block. The binary attributes are \emph{Hispanic or Latino}, \emph{White}, \emph{Black or African American}, \emph{American Indian and Alaska Native}, \emph{Asian}, \emph{Native Hawaiian and Other Pacific Islander}, \emph{Some Other Race}. The \textbf{AgeGender} dataset is extracted from the SF1 tabulation PCT12. For each of the $73,426$ census tracts in the dataset, a record has a binary gender variable and 103 possible age values. The 2010 \textbf{Brazil} dataset is obtained from IPUMS \cite{ipums} and consists of $20,635,472$ census records  from Brazil. We extract the following attributes: \emph{state} (25 possible values), \emph{occupation} (437 possible values), \emph{age} (101 possible values), and \emph{gender} (2 possible values).

\subsection{Evaluation Measures}\label{sec:evalmeasures}
We consider the nested analysis setting (Section \ref{subsec:select:nested}) in which the primary analysis is represented by $\mech_1 (\datavec)= \bmat_1 \datavec + N(\vec{0}, \covar_1)$. The secondary analysis is represented by $\mech_2 (\datavec) = \bmat_2 \datavec + N(\vec{0}, \covar_2)$. Thus, the primary analysis  can also be done, but less accurately, with the output of $\mech_2$. However, $\mech_2$ supports additional analyses that $\mech_1$ cannot.

The output $\outp_*$ of the common mechanism is used to estimate whether the output of $\mech_2$ is accurate enough to perform the primary analysis -- if at least $x\%$ of the marginal cells are believed to have an SNR at least $y$ and then the residual mechanism $\mech_2^\prime$ is run to recreate the output of $\mech_2$ without wasting privacy loss budget (otherwise the residual mechanism $\mech_1^\prime$ is used). An \textbf{alternate approach}, that doesn't use the common mechanism, is to reserve some privacy budget to estimate the SNR (by getting a coarse noisy answer to $\bmat_1 \datavec$ and then using that noisy answer to estimate SNR as in Section \ref{subsec:select:nested}) and then run either $\mech_1$ or $\mech_2$. The alternate approach uses the \emph{optimal} Gaussian mechanism \cite{xiao2020optimizing} that matches the variance of the common mechanism (for an apples-to-apples comparison) while minimizing privacy cost. We measure the following quantities:
\begin{itemize}[leftmargin=*]
\item \underline{$\rho$}: the concentrated differential privacy parameter (zcdp) \cite{zcdp} that represents the total privacy loss budget.
\item \underline{$\%\mech_1$, $\%\mech_2$}: based on the ground truth (computed from the true count divided by noise std), the percent of the time that $\mech_1$ (resp., $\mech_2$) should have been chosen. A good decision-making strategy should outperform the maximum of these two. In Section \ref{sec:exp:census}, where we must choose between 4 mechanisms, we report $\%\mech_1$, $\%\mech_2$, $\%\mech_3$, $\%\mech_4$.
\item \underline{Acc}: accuracy of the selection based on the common mechanism. This is the percentage of time that the correct mechanism has been chosen without wasting any privacy budget.
\item \underline{$\%$PLB Saved}: this is how much privacy loss budget the optimal alternate approach needs to allocate to the estimation of SNR in order to match the estimation quality of the common mechanism. This is how much privacy loss budget is saved by using the common mechanism methodology instead of the traditional alternate approach. We express this as a percentage of the total privacy loss budget. Note that $\%$PLB Saved depends only on the query matrices of $\mech_1$ and $\mech_2$, so there is just one  $\%$PLB Saved value per table.
\end{itemize}

\subsection{Marginals on HispRace}\label{sec:exp:hr}

For the \emph{HispRace} dataset, we consider two settings. In the first case, $\mech_1$ adds independent noise to 1-way marginals while $\mech_2$ adds independent noise to 2-way marginals. For each census block, the analysts must choose between $\mech_1$ and $\mech_2$. When $\mech_2$ is accurate enough to allow an analyst to derive 1-way marginals that exceed an SNR bound, then $\mech_2$ is preferred. In the second set of experiments, $\mech_1$ adds independent noise to 1-way marginals while $\mech_2$ adds independent noise to the identity query (i.e., each cell of the data vector $\datavec$). Table \ref{tab:hr-rho} shows the results as the privacy loss is varied. The common mechanism allows the analyst to correctly choose the right analysis with high accuracy. In the case of 1-way vs. 2-way marginals, the privacy loss budget saved (compared to methods that allocate some privacy loss budget for SNR estimation) is significant (\textcolor{black}{$75\%$}), while for 1-way vs. identity, the savings are more moderate (\textcolor{black}{$6.25\%$}). Tables \ref{tab:hr-y} and \ref{tab:hr-x} show the accuracy of selection based on the common mechanism as the SNR parameters $x$ (desired fraction of cells with high signal) and $y$ (desired minimum signal-to-noise ratio) are varied. 

Overall, when one wishes to run either $\mech_1$ or $\mech_2$, then the common mechanism represents information that comes for free because both mechanisms provide it. This information is accurate enough for choosing between the mechanisms and does not waste privacy budget. 

Note that by avoiding the traditional approach of reserving privacy loss budget for making a decision, another benefit is the elimination of a tuning parameter -- how much of the privacy budget to reserve for decision-making.

\begin{table}
\parbox{.49\linewidth}{
\centering
 \begin{tabular}{| c| c | c | c | }
 \hline
 \multicolumn{4}{|c|}{\textbf{1-way vs. 2-way Marginals}}\\
 \multicolumn{1}{|c}{$\rho$} &  \multicolumn{1}{c}{\asfname} & \multicolumn{1}{c}{\ascname}  & \multicolumn{1}{c|}{Acc}  \\ [0.5ex] 
  \hline
   2  & 45.54 & 54.46 & 98.64 \\   \hline
 1   & 54.35 & 45.65 & 97.98\\ \hline
 $1/2$  &  65.47& 34.53& 98.37 \\  \hline
  $1/8$   & 84.12 & 15.81 & 98.84\\   \hline
 $1/32$     & 94.41 & 5.59 & 99.56 \\ \hline
  \multicolumn{4}{|c|}{\textbf{$\%$PLB Saved}: \textcolor{black}{$75\%$}}\\\hline
\end{tabular}

}
\hfill
\parbox{.49\linewidth}{
\centering
 \begin{tabular}{| c| c | c | c | }
 \hline
 \multicolumn{4}{|c|}{\textbf{1-way vs. Identity}}\\
 \multicolumn{1}{|c}{$\rho$} &  \multicolumn{1}{c}{\asfname} & \multicolumn{1}{c}{\ascname}  & \multicolumn{1}{c|}{Acc}  \\ [0.5ex] 
  \hline
  2  & 50.20 & 49.80 &  95.00 \\\hline
    1  & 61.46 & 38.54 & 95.37 \\  \hline
 $1/2$  & 71.32 & 28.68 & 95.40  \\   \hline
  $1/8$  & 88.27 & 11.73 & 97.53 \\ \hline
 $1/32$   & 95.98 & 4.02 & 99.08   \\
 \hline
 \multicolumn{4}{|c|}{\textbf{$\%$PLB Saved}: \textcolor{black}{$6.25\%$}}\\\hline
\end{tabular}
}
\caption{Experiments on \textbf{HispRace} dataset as zCDP privacy budget $\rho$ varies. The SNR parameters are $(x,y)=(0.5, 5)$. }
\label{tab:hr-rho}
\end{table}

\begin{table}
\parbox{.49\linewidth}{
\centering
 \begin{tabular}{| c | c | c | c |  }
 \hline
 \multicolumn{4}{|c|}{\textbf{1-way vs. 2-way Marginals}}\\
 \multicolumn{1}{|c}{$y$} &  \multicolumn{1}{c}{\asfname} & \multicolumn{1}{c}{\ascname}  & \multicolumn{1}{c|}{Acc}  \\ [0.5ex] 
  \hline
   2 & 58.08 & 41.92 & 95.56 \\
 \hline
 3 & 70.57 & 29.43 & 97.22 \\
 \hline
  4 & 78.75 & 21.25 & 98.20 \\
 \hline
  5 & 84.19 & 15.81 & 98.83 \\
 \hline
  6 & 87.82 & 12.18 & 99.22\\
 \hline
  8 & 92.10 & 7.90 & 99.61 \\
 \hline
 \multicolumn{4}{|c|}{\textbf{$\%$PLB Saved}: \textcolor{black}{$75\%$}}\\\hline
\end{tabular}
}
\hfill
\parbox{.49\linewidth}{
\centering
 \begin{tabular}{| c | c | c | c |  }
 \hline
 \multicolumn{4}{|c|}{\textbf{1-way vs. Identity}}\\
 \multicolumn{1}{|c}{$y$} &  \multicolumn{1}{c}{\asfname} & \multicolumn{1}{c}{\ascname}  & \multicolumn{1}{c|}{Acc}  \\ [0.5ex] 
  \hline
  2  & 64.52 & 35.48 & 88.83 \\
 \hline
 3  & 76.58 & 23.42 & 93.56 \\
 \hline
  4  & 83.85 & 16.15 & 96.08 \\
 \hline
  5  & 88.27 & 11.73 & 97.54 \\
 \hline
   6  & 91.07 & 8.93 & 98.35 \\
 \hline
  8 & 94.27 & 5.73 & 99.17 \\
 \hline
 \multicolumn{4}{|c|}{\textbf{$\%$PLB Saved}: \textcolor{black}{$6.25\%$}}\\\hline
\end{tabular}
}
\caption{Experiments on \textbf{HispRace} dataset for zCDP parameter $\rho=1/8$ and SNR parameters $(0.5, y)$ as $y$ varies.}
\label{tab:hr-y}
\end{table}

\begin{table}
\parbox{.49\linewidth}{
\centering
 \begin{tabular}{| c | c | c | c |  }
 \hline
 \multicolumn{4}{|c|}{\textbf{1-way vs. 2-way Marginals}}\\
 \multicolumn{1}{|c}{$x$} &  \multicolumn{1}{c}{\asfname} & \multicolumn{1}{c}{\ascname}  & \multicolumn{1}{c|}{Acc}  \\ [0.5ex] 
  \hline
   0.2   &79.61  &20.39  & 99.07\\
 \hline
 0.3  & 81.25 & 18.75 & 98.97 \\
 \hline
  0.4  & 82.96  & 17.04 & 98.95 \\
 \hline
  0.5  & 84.19 & 15.81 & 98.83 \\
  \hline
  0.6 & 97.95 & 2.05 & 99.85 \\
 \hline
 \multicolumn{4}{|c|}{\textbf{$\%$PLB Saved}: \textcolor{black}{$75\%$}}\\\hline
\end{tabular}
}
\hfill
\parbox{.49\linewidth}{
\centering
 \begin{tabular}{| c | c | c | c |  }
 \hline
 \multicolumn{4}{|c|}{\textbf{1-way vs. Identity}}\\
 \multicolumn{1}{|c}{$x$} &  \multicolumn{1}{c}{\asfname} & \multicolumn{1}{c}{\ascname}  & \multicolumn{1}{c|}{Acc}  \\ [0.5ex] 
  \hline
   0.2   & 84.42 & 15.58 & 97.05\\
 \hline
 0.3  & 85.86 & 14.14 & 97.44\\
 \hline
  0.4  & 87.30 & 12.70 & 97.55 \\
 \hline
  0.5  & 88.27 & 11.73 & 97.52\\
 \hline
   0.6  & 98.50 & 1.50 & 99.65\\
 \hline
 \multicolumn{4}{|c|}{\textbf{$\%$PLB Saved}: \textcolor{black}{$6.25\%$}}\\\hline
\end{tabular}
}
\hfill
\caption{Experiments on \textbf{HispRace} dataset for zCDP parameter $\rho=1/8$ and SNR parameters $(x, 5)$ as $x$ varies.}
\label{tab:hr-x}
\end{table}

\subsection{Marginals on the Brazil Dataset}\label{sec:exp:br}
In the \textbf{Brazil} dataset, we consider the setting where, for each combination of \emph{state} and \emph{occupation}, the analyst needs to choose whether to run $\mech_1$ to produce noisy 1-way marginals or $\mech_2$ to produce noisy 2-way marginals by adding independent noise.

The accuracy of making the choice, for each state/occupation combination is shown in Table \ref{tab:brazil-rho}. Again, the common mechanism provides enough information for choosing between the two mechanisms (i.e. choosing between which residual mechanism to run). It also provides significant privacy budget  savings (\textcolor{black}{50.5\%}) compared to the approach that first allocates privacy loss budget to estimating SNR just as accurately, before making a choice, with the side benefit being that there is no privacy budget allocation tuning parameter necessary when using the common mechanism. The accuracy of selection based on the common mechanism, as we vary the SNR parameters, is shown in Table \ref{tab:brazil-xy} and again shows fairly good accuracy.

\begin{table}
\parbox{.49\linewidth}{
\centering
  \begin{tabular}{| c| c | c | c | }
 \hline
 \multicolumn{4}{|c|}{\textbf{1-way vs. 2-way Marginals}}\\
 \multicolumn{1}{|c}{$\rho$} &  \multicolumn{1}{c}{\asfname} & \multicolumn{1}{c}{\ascname}  & \multicolumn{1}{c|}{Acc}  \\ [0.5ex] 
  \hline
 2 &  71.47 & 28.53  & 96.95 \\
   \hline
    1  & 71.47  &  28.53 & 97.86\\
   \hline
 $1/2$  & 78.65  & 21.35  & 98.29 \\
   \hline
 $1/8$    & 85.30  & 14.70  & 98.94 \\
   \hline
 $1/32$    & 90.66  & 9.34  & 99.20 \\
 \hline
     \multicolumn{4}{|c|}{\textbf{$\%$PLB Saved}: \textcolor{black}{$50.5\%$}}\\\hline
\end{tabular}
}
\caption{Experiments on \textbf{Brazil} dataset as zCDP privacy budget $\rho$ varies. The SNR parameters are  $(x,y)=(0.3, 3)$.}
\label{tab:brazil-rho}
\end{table}

\begin{table}
\parbox{.49\linewidth}{
\centering
 \begin{tabular}{| c | c | c | c |  }
 \hline 
  \multicolumn{4}{|c|}{\textbf{1-way vs. 2-way Marginals}}\\
 \multicolumn{1}{|c}{$x$} &  \multicolumn{1}{c}{\asfname} & \multicolumn{1}{c}{\ascname}  & \multicolumn{1}{c|}{Acc}  \\ [0.5ex] 
  \hline
   0.2  &  58.56 &  41.44  & 96.27\\
 \hline
 0.3  & 71.47  & 38.53  &  96.95\\
 \hline
  0.4  &  81.14 & 18.86  & 97.55 \\ 
 \hline
  0.5  &  88.78 & 11.22  & 98.46\\
 \hline
  0.6 &  94.58 & 5.42  & 99.18 \\
 \hline
      \multicolumn{4}{|c|}{\textbf{$\%$PLB Saved}: \textcolor{black}{$50.5\%$}}\\\hline
\end{tabular}
}
\hfill
\parbox{.49\linewidth}{
\begin{tabular}{| c | c | c | c |  }
\hline
 \multicolumn{4}{|c|}{\textbf{1-way vs. 2-way Marginals}}\\
 \multicolumn{1}{|c}{$y$} &  \multicolumn{1}{c}{\asfname} & \multicolumn{1}{c}{\ascname}  & \multicolumn{1}{c|}{Acc}  \\ [0.5ex] 
  \hline
 3  &  71.47 & 28.53  & 96.95 \\
 \hline
  4  & 71.47  & 28.53  & 98.41 \\
 \hline
  5 & 75.78  & 24.22  & 99.03 \\
\hline
   6  & 78.65  & 21.35  & 99.00 \\
 \hline
   8 & 80.94  & 19.06  &99.40 \\
 \hline
      \multicolumn{4}{|c|}{\textbf{$\%$PLB Saved}: \textcolor{black}{$50.5\%$}}\\\hline
\end{tabular}
}
\caption{Experiment on \textbf{Brazil} dataset for zCDP parameter $\rho=2$. Left: SNR parameters $(x, 3)$ as $x$ varies. Right: SNR parameters $(0.3, y)$ as $y$ varies.}
\label{tab:brazil-xy}
\end{table}

\subsection{Census Application: Age/Gender Histograms}\label{sec:exp:census}

Our next set of experiments is a case study for a data product that will be released as part of the 2020 Decennial Census Detailed Demographics and Housing Characteristics \cite{dhc}.

\subsubsection{Problem Description}
The goal of this data product is to produce age-by-gender histograms for different sub-populations, such as for an ethnic group in a given region. Since the sub-population might be sparse, one of 4 pre-defined age bucketization schemes will be used \cite{dhc}: 
\begin{itemize}[leftmargin=*]
\item \textbf{Total}. This consists of one age bucket: $[0,103)$. In this case, the age-by-gender histogram is simply the number of females and the number of males.
\item \textbf{Age4}. This consists of the following four buckets: $[0,18)$; $[18,45)$; $[45, 65)$; $[65,103]$.
\item \textbf{Age9}. This consists of the following nine buckets: $[0,5)$; $[5, 18)$; $[18,25)$; $[25,35)$; $[35,45)$; $[45,55)$; $[55,65)$; $[65, 75)$; $[75,103]$.
\item\textbf{Age23}. This consists of the following 23 buckets: $[0,5)$; $[5 10)$; $[10,15)$; $[15,18)$; $[18,20)$; $[20,21)$; $[21,22)$; $[22,25)$; $[25,30)$; $[30,35)$; $[35,40)$; $[40,45)$; $[45,50)$; $[50, 55)$; $[55,60)$; $[60,62)$; $[62,65)$; $[65,67)$; $[67,70)$; $[70,75)$; $[75,80)$; $[80,85)$; $[85,103]$.
\end{itemize}
We note that these represent nested analyses, as \textbf{Age23} is a refinement of \textbf{Age9}, which is a refinement of \textbf{Age4}, which is a refinement of \textbf{Total}. The idea is that the smaller the sub-population is, the coarser the age buckets should be in order for the noise not to overwhelm the actual counts.

\subsubsection{The Census  algorithm}
One of the algorithms we compare against is the DHC algorithm, which is the one that will actually be used for the problem \cite{dhc} by the Census Bureau. This DHC algorithm will use a fraction $\gamma$  of the privacy loss budget to estimate the total size of the sub-population. There are also three threshold parameters $\theta_1 < \theta_2 < \theta_3$. If the noisy sub-population count is $<\theta_1$, the remaining privacy budget will be used to produce the gender by age histogram using the \textbf{total} bucketization. If the noisy sub-population count is in the range $[\theta_1,\theta_2)$, then the \textbf{Age4} bucketization will be used. If it is in the range of $[\theta_2,\theta_3)$, then \textbf{Age9} will be used. Otherwise, \textbf{Age23} will be used.

Thus the algorithm has 4 parameters that must be carefully tuned: $\gamma, \theta_1,\theta_2,\theta_3$. The decision on parameter values has not been made  public (possibly a difficult decision) but the rest of the algorithm is public.

\subsubsection{The Common Mechanism Approach}
In the problem setting, there are four algorithms to consider: $\mech_1,\mech_2,\mech_3,\mech_4$, which produce noisy gender by age histograms with the bucketizations \textbf{total}, \textbf{Age4}, \textbf{Age9}, \textbf{Age23}, respectively, by adding independent Gaussian noise. We define 3 common mechanisms: $CM_{1234}$ is common to $\mech_1,\mech_2,\mech_3,\mech_4$ (i.e., it is the maximal mechanism answerable from all of them). It is obtained by numerically solving the optimization in Equation \ref{prob:min-trace} with a constraint added for each mechanism. $CM_{234}$ is common to $\mech_2, \mech_3, \mech_4$, and $CM_{34}$ is common to $\mech_3$ and $\mech_4$. It turns out that $CM_{1234}$ is answerable from $CM_{234}$, which is answerable from $CM_{34}$, therefore allows us to perform the following procedure that wastes no privacy loss budget:
\begin{itemize}[leftmargin=*]
\item Get the output $\outp_{1234}$ of $\mech_{1234}$ and decide whether the \textbf{total} bucketization should be used based on SNR. If yes, run the residual mechanism $RM_1$ such that $(CM_{1234}, RM_1)$ is equivalent to $\mech_1$.
\item If no, run the residual mechanism $RM_{234}$ such that $(\mech_{1234}, RM_{234})$ is equivalent to $CM_{234}$. Use that to decide whether to use $\textbf{Age4}$. If yes, run the residual mechanism $RM_2$ such that $(CM_{234}, RM_2)$ is equivalent to $\mech_2$.
\item If no, run the residual mechanism $RM_{34}$ such that $(\mech_{234}, RM_{34})$ is equivalent to $CM_{34}$. Use that to decide whether to use $\textbf{Age9}$. If yes, run the residual mechanism $RM_{3}$ such that $(CM_{34}, RM_3)$ is equivalent to $\mech_3$.
\item If no, run the residual mechanism $RM_4$ such that $(CM_{34}, RM_4)$ is equivalent to $\mech_4$.
\end{itemize}
Thus, at the end, one gets something that is equivalent to either $\mech_1,\mech_2,\mech_3$ or $\mech_4$ without wasting any privacy budget.

\subsubsection{The Alternative Approach.}\label{subsec:ag:alternate} We also consider a third approach that mirrors the previous experiments. Instead of making decisions based on the common mechanism, at each step, some privacy loss budget is allocated to obtain the best linear Gaussian mechanism to estimate the SNR of the histogram being considered. This is also an alternative to making decisions based on noisy population counts.

\subsubsection{Results}
Since the input data that the DHC algorithm will use is not public, we use the \textbf{AgeGender} dataset described in Section \ref{sec:data} to produce an Age by Gender histogram at each census tract using the three algorithms described. 

Since the DHC algorithm requires additional parameters, we give it a strong non-private advantage: the $\theta_1,\theta_2,\theta_3$ are learned using a non-private logistic regression model, and the DHC algorithm is given the exact sub-population count (i.e., a noiseless threshold).

We set the desired SNR parameters to be $x=0.5$ and $y=20$. In a histogram cell, a ratio of 20 between a count $c$ and the privacy noise standard deviation $\sigma$ means that $95\%$ of the time, the relative error of the noisy cell count  is at most $2\sigma/c=2/20=10\%$.

\begin{figure}
    \centering
    \includegraphics[width=0.4\textwidth,clip=true,trim=0.0cm 0.0cm 0cm 1.3cm]{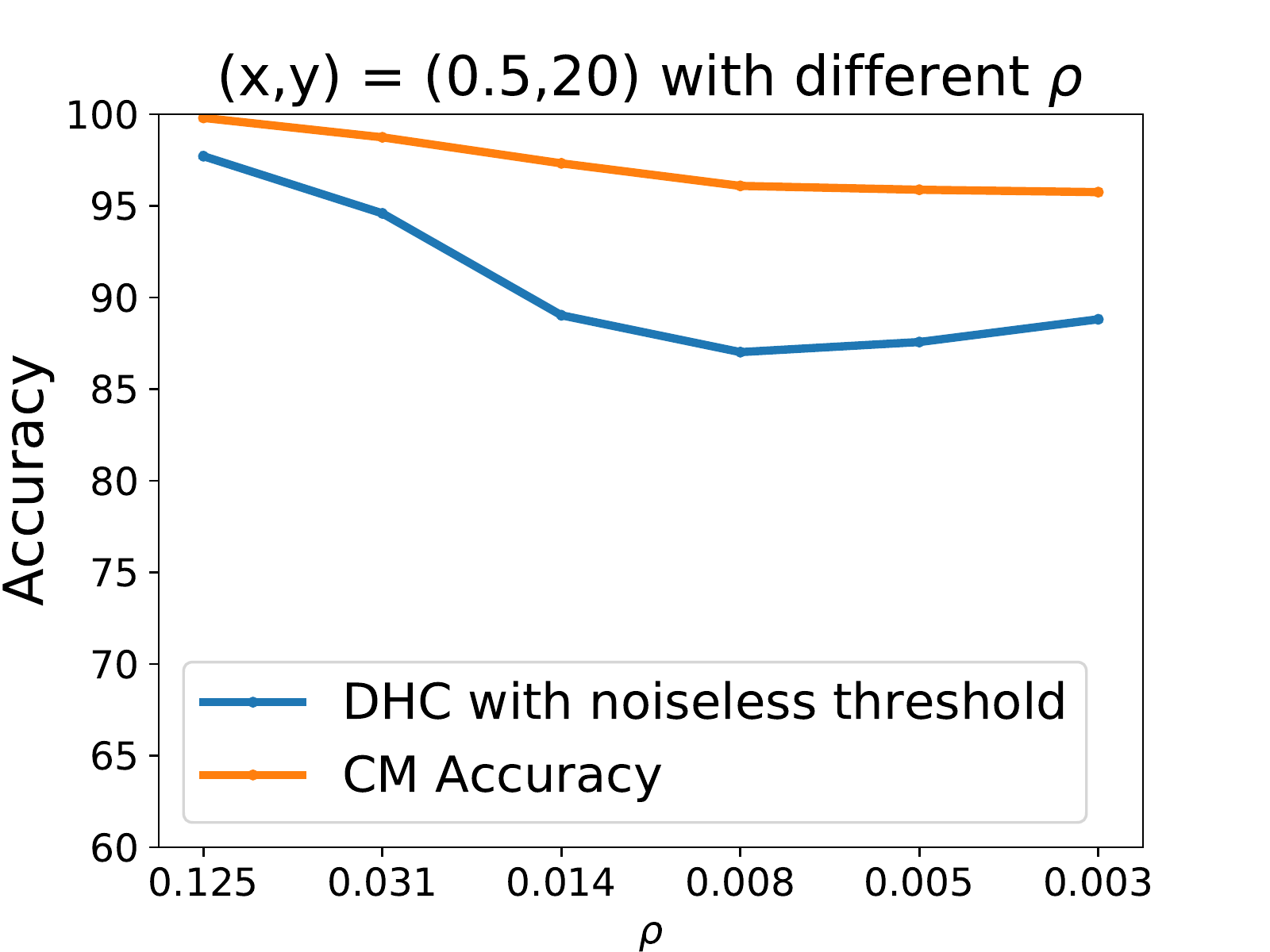}
\caption{Common mechanism vs. tuned DHC algorithm as privacy budget  $\rho$ varies.}
    \label{fig:agegender-s}
\end{figure}

In Figure \ref{fig:agegender-s}, we compare the accuracy of the DHC algorithm to the common mechanism in choosing the right age bucketization to use. Even with the advantages we gave it (e.g., tuned parameters and noise-free thresholds), it is still outperformed by the common mechanism, which avoids all those tuning parameters. This shows that the noisy information provided by the common mechanism for free is more informative than the population thresholds, even when the population thresholds are completely accurate.

We next compare the common mechanism to the alternate approach (Section \ref{subsec:ag:alternate}) that, for an apples-to-apples comparison, allocates privacy budget to the optimal linear Gaussian mechanism that has equal variance to the common mechanism, but with smallest possible privacy cost. Table \ref{tab:ag-alt} shows the accuracy of choosing the correct age bucketization and the fraction of PLB saved by using the common mechanism (nearly $50\%$). Since the choice among the four bucketizations is done sequentially, the sooner one is chosen, the better for the alternate approach since it no longer has to allocate PLB for future choices. For this reason, the percentage of PLB saved varies with the experimental setting.

\begin{table}
\centering
 \begin{tabular}{| c  | c | c | c | c | c| c |
 }
 \hline
 $\rho$  &  $\%\mech_1$ & $\%\mech_2$ & $\%\mech_3$ &$\%\mech_4$  & Acc  &\makecell{\textbf{\% PLB} \\ \textbf{saved}} \\ [0.5ex] 
  \hline
 $1/8$    & 0.34 & 0.68 & 1.64 & 97.34 & 99.80 & 49.67 \\
   \hline
 $1/32$  & 0.39 & 1.31 &12.61 &85.69 & 98.74 & 49.59 \\
 \hline
  $1/72$ & 0.45 & 3.46 &31.51 &64.58 & 97.32 & 48.92 \\
 \hline
 $1/128$    & 0.50 & 8.08 & 49.95&41.47 & 96.09 & 47.74\\
 \hline
  $1/200$  & 0.56 & 14.50 & 61.43 &23.51 & 95.88 & 46.11\\
 \hline
  $1/288$  & 0.63 & 22.55 & 64.85 & 11.97 & 95.75  & 46.06\\
 \hline
\end{tabular}
 \caption{Privacy budget savings vs. the alternate approach.}
\label{tab:ag-alt}
\end{table}

\section{Conclusions}\label{sec:conc}
 In this paper, we formalized the problem and provided algorithms for the computation of the common mechanism $\commech$ for two linear Gaussian mechanisms $\mech_1$ and $\mech_2$. The common mechanism represents information that is provided by both mechanisms, while the residual mechanisms $\mech^\prime_1$ and $\mech^\prime_2$ reflect the remaining information in $\mech_1$ and $\mech_2$ after the information from $\commech$ has been removed from them. We presented an application where an analyst can decide whether to get answers of $\mech_1$ or $\mech_2$ using the help of the common and residual mechanisms. This represents another tool that can be used for differentially private algorithm design.

\section*{Acknowledgments}\label{sec:ack}
This work was supported by NSF Awards CNS-1702760 and CNS-1931686.

\bibliographystyle{ACM-Reference-Format}
\bibliography{ref.bib}

\ifbool{ARXIV}{
\clearpage
\appendix

\section{Proofs from Section \ref{sec:problemdef}}
\printProofs[section3]
\section{Proofs from Section \ref{sec:common}}
\printProofs[section5]

}{}
\end{document}